\documentstyle[eqsecnum,preprint,aps,epsfig,floats]{revtex}

\tighten

\newcommand{\req}[1]{(\ref{#1})}

\newcommand{\Beq}{\begin{equation}}
\newcommand{\Eeq}{\end{equation}}
\newcommand{\beq}{\begin{displaymath}}
\newcommand{\eeq}{\end{displaymath}}
\newcommand{\Beqa}{\begin{eqnarray}}
\newcommand{\Eeqa}{\end{eqnarray}}
\newcommand{\beqa}{\begin{eqnarray*}}
\newcommand{\eeqa}{\end{eqnarray*}}
\newcommand{\Bml}{\begin{mathletters}}
\newcommand{\Eml}{\end{mathletters}}

\newcommand{\nUV}{\mu_{R}^{2}}
\newcommand{\nIR}{\mu_{F}^{2}}
\newcommand{\nO}{\mu_{0}^{2}}

\newcommand{\x}{\overline{x} \,}
\newcommand{\y}{\overline{y} \,}
\newcommand{\uu}{\overline{u} \,}
\newcommand{\vv}{\overline{v} \,}
\newcommand{\eIR}{\epsilon_{IR}}
\newcommand{\eUV}{\epsilon_{UV}}
\newcommand{\Eir}{\hat{\epsilon}_{IR}}
\newcommand{\Euv}{\hat{\epsilon}_{UV}}
\newcommand{\UV}{\eta_{UV}}
\newcommand{\IR}{\eta_{IR}}
\newcommand{\ir}{\tilde{\eta}_{IR}}
\newcommand{\Nuv}{4 \pi \mu_{UV}^{2}}
\newcommand{\Nir}{4 \pi \mu_{IR}^{2}}
\newcommand{\Li}{\mbox{Li}_{2}}

\begin{document}

\preprint{IRB-TH-4/97}

\draft

\title{Complete next-to-leading order perturbative
       QCD prediction for the pion electromagnetic form factor}

\author{B. Meli\'{c},
        B. Ni\v{z}i\'{c},
        and K. Passek\thanks{Electronic addresses: melic@thphys.irb.hr,
                             nizic@thphys.irb.hr, passek@thphys.irb.hr}}

\address{Theoretical Physics Division, Rudjer Bo\v{s}kovi\'{c} Institute, \\
        P.O. Box 1016, HR-10001 Zagreb, Croatia}

\date{March 1999}

\maketitle

\begin{abstract}
We present the results of a complete leading-twist
next-to-leading order (NLO) QCD analysis of the 
spacelike pion electromagnetic form factor at large momentum
transfer $Q$.
We have studied their dependence on the form of 
the pion distribution amplitude.  
For a given distribution amplitude, 
we have examined the sensitivity of the predictions 
to the choice of the renormalization and factorization scales.
Theoretical uncertainty of the LO results related to the renormalization
scale ambiguity has been significantly reduced by including the
NLO corrections.
Adopting the criteria according to which a NLO prediction
is considered reliable if, both, the ratio of the NLO to LO contributions
and the strong coupling constant are reasonably small,
we find that reliable perturbative predictions 
for the pion electromagnetic form factor with 
all distribution amplitudes considered can already be made at a momentum
transfer $Q<10$ GeV, with corrections to the LO
results being typically of the order of $\sim 20\%$.
To check our predictions and to discriminate 
between the distribution amplitudes, it is necessary to obtain 
experimental data extending to higher values of $Q$.
\end{abstract}


\narrowtext

\section{Introduction}

Exclusive processes involving large momentum transfer 
are among the most interesting and challenging tests
of quantum chromodynamics (QCD).

The framework for analyzing such processes within the context
of perturbative QCD (PQCD)
has been developed by Brodsky and Lepage
\cite{BrL79etc}, Efremov and Radyushkin \cite{EfR80},
and Duncan and Mueller \cite{DuM80} 
(see Ref. \cite{BrL89etc} for reviews).
They have demonstrated, to all orders in perturbation theory,
 that exclusive amplitudes involving
large momentum transfer factorize into a convolution
of a process-independent and perturbatively incalculable
distribution amplitude, one for each hadron involved in the amplitude, with
a process-dependent and perturbatively calculable 
hard-scattering amplitude.

Within the framework developed in Refs. \cite{BrL79etc,EfR80,DuM80},
leading-order (LO) predictions have been obtained for many 
exclusive processes.
It is well known, however, that, unlike in QED, the LO predictions in PQCD 
do not have much predictive power, and that higher-order corrections
are essential for many reasons.
In general, they have a stabilizing effect reducing the dependence of the 
predictions on the schemes and scales. Therefore, to achieve a complete 
confrontation between theoretical predictions and experimental data, it is
very important to know the size of radiative corrections to the LO
predictions.
The list of exclusive processes at large momentum transfer analyzed at 
next-to-leading order (NLO) is very short and includes only three processes:
the pion electromagnetic form factor \cite{FiG81,DiR81,Sa82,KhaR85,BraT87,KaM86},
the pion transition form factor \cite{KaM86,AgC81,Bra83},
and photon-photon annihilation into two flavor-nonsinglet helicity-zero
mesons, $\gamma \gamma \rightarrow M \overline{M}$ ($M=\pi$, $K$) \cite{Ni87}.

\widetext

In leading twist, the pion electromagnetic form factor
(the simplest exclusive quantity) can be written as
\Beq
    F_{\pi}(Q^{2})=\int_{0}^{1} dx \int_{0}^{1} dy \;
                    {\sl \Phi}^{*}(y,\nIR) \;
                          T_{H}(x,y,Q^{2},\nUV,\nIR) \;
                    {\sl \Phi}(x,\nIR) \, .
\label{eq:piffcf}
\Eeq

\narrowtext

Here ${\sl \Phi}(x,\nIR)$ is the pion distribution amplitude, i.e.,
the probability amplitude for finding the valence
$q_1 \overline{q}_2$ Fock state in the initial pion with the 
constituents carrying the longitudinal momentum $x P$ and $(1-x) P$; 
$T_{H}(x,y,Q^{2},\nUV,\nIR)$ is the hard-scattering amplitude, i.e.,
the amplitude for a parallel 
$q_1 \overline{q}_2$ pair of the total momentum $P$ hit by a virtual photon
$\gamma^*$ of momentum $q$ to end up as a parallel
$q_1 \overline{q}_2$ pair of momentum $P'=P+q$;
${\sl \Phi}^{*}(y,\nIR)$ is the amplitude for the final state
$q_1 \overline{q}_2$ to fuse back into a pion;
$Q^2=-q^2$ is the momentum transfer in the process and is supposed to be
large; $\mu_R$ is the renormalization (or coupling constant) scale
and $\mu_F$ is the factorization (or separation) scale at which
soft and hard physics factorize.

The hard-scattering amplitude $T_{H}$ can be calculated in 
perturbation theory and represented as a
series in the QCD running coupling constant $\alpha_S(\nUV)$.
The function ${\sl \Phi}$ is intrinsically nonperturbative, 
but its evolution can be calculated perturbatively.

Although the PQCD approach of Refs. \cite{BrL79etc,EfR80,DuM80}
undoubtedly represents an adequate and efficient tool for analyzing
exclusive processes at very large momentum transfer,
its applicability to these processes at experimentally accessible 
momentum transfer has long been debated and attracted much attention.
The concern has been raised \cite{IsLS84etc,Rad91} that, 
even at very large momentum transfer,
important contributions to these processes can arise from nonfactorizing 
end-point contributions of the distribution amplitudes with $x \sim 1$.
It has been shown, however, that the incorporation of the Sudakov suppression
effectively eliminates these soft contributions and that the PQCD approach 
to the pion form factor begins to be self-consistent for a momentum transfer
of about $Q^2 > 4$ GeV$^2$ \cite{LiS92etc} 
(see also Ref. \cite{JaK93}).

To obtain the complete NLO prediction
for the pion form factor requires calculating NLO corrections to both the
hard-scattering amplitude and the evolution kernel 
for the pion distribution amplitude.

The NLO predictions for the pion form factor obtained in
Refs.  \cite{FiG81,DiR81,Sa82,KhaR85,BraT87} are incomplete in so far as
only the NLO correction to the hard-scattering amplitude has been
considered, whereas the corresponding NLO corrections to the evolution
of the pion distribution amplitude have been ignored.
Apart from not being complete, the results of the calculations presented in
Refs.  \cite{FiG81,DiR81,Sa82,KhaR85,BraT87} do not agree with one another.

Evolution of the distribution amplitude can be obtained
by solving the differential-integral evolution
equation using the moment method.
In order to determine the NLO corrections to the
evolution of the distribution amplitude,
it is necessary to calculate two-loop corrections to the evolution
kernel. 
These have been computed by different authors and the obtained results
are in agreement \cite{DiR84etc}.
Because of the complicated structure of these corrections, it
is possible to obtain numerically only the first few moments of the evolution
kernel \cite{MiR86}.
Using these incomplete results, the first attempt 
to include the NLO corrections to the evolution
of the distribution amplitude in the NLO analysis of the
pion form factor was obtained in Ref. \cite{KaM86}. 
It has been found that the NLO corrections to the evolution of the pion
distribution amplitude as well as to the pion form factor are tiny.

Considerable progress has recently been made in understanding
the NLO evolution of the pion distribution amplitude \cite{Mu94etc}.
Using conformal constraints, the complete formal solution of the NLO
evolution equation has been obtained.
Based on this result, it has been found that, contrary to the estimates
given in Ref. \cite{KaM86}, the NLO corrections to the evolution
of the distribution amplitude are rather large.  
It has been concluded that because of the size of the discovered
corrections, and their dependence upon the input distribution amplitude,
the evolution of the distribution amplitude has to be included
in the NLO analysis of exclusive processes at large momentum transfer.

The purpose of this paper is to present a complete leading-twist NLO 
QCD analysis
of the spacelike pion electromagnetic form factor at large momentum transfer.

The plan of the paper is as follows.
To check and verify the results obtained in
Ref. \cite{FiG81,DiR81,Sa82,KhaR85,BraT87}, in Sec. II we
carefully calculate all one-loop diagrams contributing to
the NLO hard-scattering amplitude for the pion form factor.
We use the Feynman gauge, the dimensional regularization method, and 
the modified minimal-subtraction ($\overline{MS}$) scheme.
Our results are in agreement with those obtained in 
Refs. \cite{FiG81} 
(modulo typographical errors listed in \cite{BraT87})
and \cite{Sa82}.
Making use of the method introduced 
in Ref. \cite{Mu94etc}, in Sec. III we determine NLO 
evolutional corrections to four available candidate pion
distribution amplitudes. 
In Sec. IV we discuss several possible choices of the renormalization
scale $\mu_R$ and the factorization scale $\mu_F$. 
In Sec. V we obtain complete NLO numerical predictions
for the pion form factor using the four candidate pion distribution
amplitudes.
For a given distribution amplitude we examine the sensitivity
of the predictions on the renormalization 
and factorization scales, $\mu_R$ and $\mu_F$, respectively.
We take $\Lambda_{\overline{MS}} = 0.2$ GeV for the calculation 
presented here. 
Section VI is devoted to discussions and some concluding remarks.

\section{NLO correction to the hard-scattering amplitude}

\begin{figure}
  \centerline{\epsfig{file=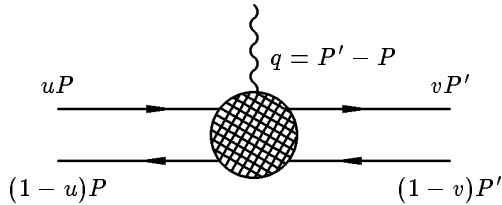,height=3.8cm,width=7.5cm,silent=}}
 \caption{Feynman diagrams describing the $(q_1 \overline{q}_2) + 
 \gamma^* \rightarrow (q_1 \overline{q}_2)$ transition 
 amplitude in terms of which 
 the hard-scattering amplitude for the pion form factor is obtained.}
 \label{f:piff}
\end{figure}
In this section we recalculate the NLO correction to the 
hard-scattering amplitude for the pion form factor.

In leading twist, the hard-scattering amplitude is obtained by evaluating  
the $(q_1 \overline{q}_2) + \gamma^* \rightarrow (q_1 \overline{q}_2)$
amplitude, which is described by the Feynman diagrams in Fig. \ref{f:piff},
with massless valence quarks collinear 
with parent mesons.  In this figure, 
$u P$ and $(1-u) P$ ($v P'$ and $(1-v) P'$)
denote the longitudinal momenta of the pion constituents before the
subtraction of collinear singularities.
In this evaluation, terms of order $m^2/Q^2$ are not included
and, since the constituents are constrained to be collinear,
terms of order $k_{\perp}^2/Q^2$ 
($k_{\perp}$ is the average transverse momentum in the meson) 
are not taken into account either.
By projecting the $q_1\overline{q}_2$ pair into a color-singlet 
pseudoscalar state the amplitude corresponding to any of the diagrams 
in Fig. \ref{f:piff} can be written in terms of a trace of a fermion loop.

By definition, the hard-scattering amplitude $T_H$ is free of collinear 
singularities and has
a well-defined expansion in $\alpha_S(\nUV)$ of the form
\Beqa
  \lefteqn{T_{H}(x,y,Q^2,\nUV,\nIR)=
   \alpha_{S}(\nUV) \, T_{H}^{(0)}(x,y,Q^2)}
            \nonumber \\ & & \quad \qquad \times \;
   \left[ 1 + \frac{\alpha_{S}(\nUV)}{\pi} \, 
	       T_{H}^{(1)}(x,y,\nUV/Q^2,\nIR/Q^2) 
                + \cdots \right] \, ,
\label{eq:TH}
\Eeqa
where
\Beq
    \alpha_{S}(\nUV)=\frac{4 \pi}{\beta_{0} \ln (\nUV/\Lambda_{QCD}^2)}
           \, ,
\label{eq:alphas}
\Eeq
and
\Beq
     \beta_{0}=11-\frac{2}{3} n_{f} \, , 
\label{eq:beta0}
\Eeq
with $n_f$ being the effective number of quark flavors and $\Lambda_{QCD}$
is the fundamental QCD parameter.

\begin{figure}
  \centerline{\epsfig{file=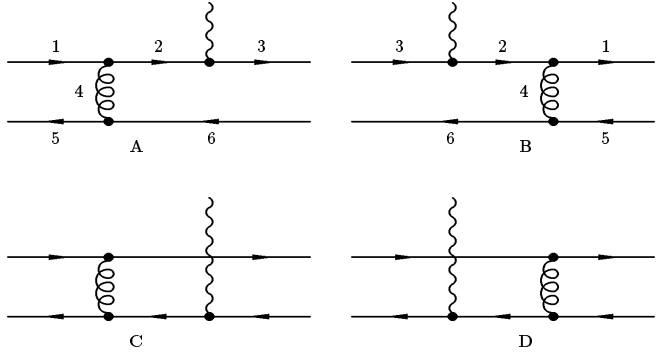,height=5cm,width=9cm,silent=}}
 \caption{Lowest-order Feynman diagrams contributing to the
      $(q_1 \overline{q}_2) +  \gamma^* \rightarrow (q_1 \overline{q}_2)$
      amplitude.}
 \label{f:loan}
\end{figure}
In the LO approximation (Born approximation)
there are only four Feynman diagrams contributing to the
$(q_1 \overline{q}_2) + \gamma^* \rightarrow (q_1 \overline{q}_2)$
transition amplitude. They are shown in Fig. \ref{f:loan}.
Evaluating these diagrams, one finds that the LO hard-scattering amplitude 
is given by

\Beq
    T_{H}^{(0)}(x,y,Q^2)=\frac{4}{3}
       \frac{16 \pi}{Q^2 (1-x) (1-y)} \, .
\label{eq:TH0}
\Eeq

To obtain this result, it is necessary to evaluate only one of the four 
diagrams. Namely, knowing the contribution of diagram $A$, the contribution 
of diagrams $B$, $C$, and $D$ can be obtained by 
making use of the isospin symmetry and time-reversal symmetry. 
In the leading-twist approximation, the isospin symmetry is exact.  
A consequence of this is that the contributions of diagrams
$A$ and $B$ are related to the contributions of $C$ and $D$, respectively.
On the other hand, diagrams $A$ and $C$ are by time-reversal symmetry
related to diagrams $B$ and $D$, respectively.

At NLO there are altogether 62 one-loop Feynman diagrams 
contributing to the 
$(q_1 \overline{q}_2) + \gamma^* \rightarrow (q_1 \overline{q}_2)$
amplitude.
They can be generated by inserting an internal gluon line into 
the leading-order diagrams of Fig. 2.
Use of the above mentioned symmetries (isospin and time-reversal) cuts 
the number of  independent one-loop diagrams to be evaluated from 
62 to 17. They are all generated from the LO diagram $A$. 
We use the notation where $Aij$ is the diagram obtained from diagram $A$
by inserting the gluon line connecting the lines $i$ and $j$,
where $i,j=1,2,\cdots,6$.
They are shown in Fig. 3 
with the exception of $A33$, $A55$, and $A66$, which give obviously
the same contribution as $A11$.
These diagrams contain ultraviolet (UV) singularities
and, owing to the fact that 
initial- and final-state quarks are massless and onshell,
they also contain 
both infrared (IR) and collinear singularities.
We use dimensional regularization in
$D=4-2\epsilon$ dimensions to regularize all three types of
singularities, distinguishing the poles $1/\epsilon$ by the
subscripts UV and IR
($D=4-2\eUV=4+2\eIR$).
Soft singularity is always accompanied by two collinear singularities
and, consequently, when dimensionally regularized, leads to the
double pole $1/\eIR^2$. 

\begin{figure}
  \centerline{\epsfig{file=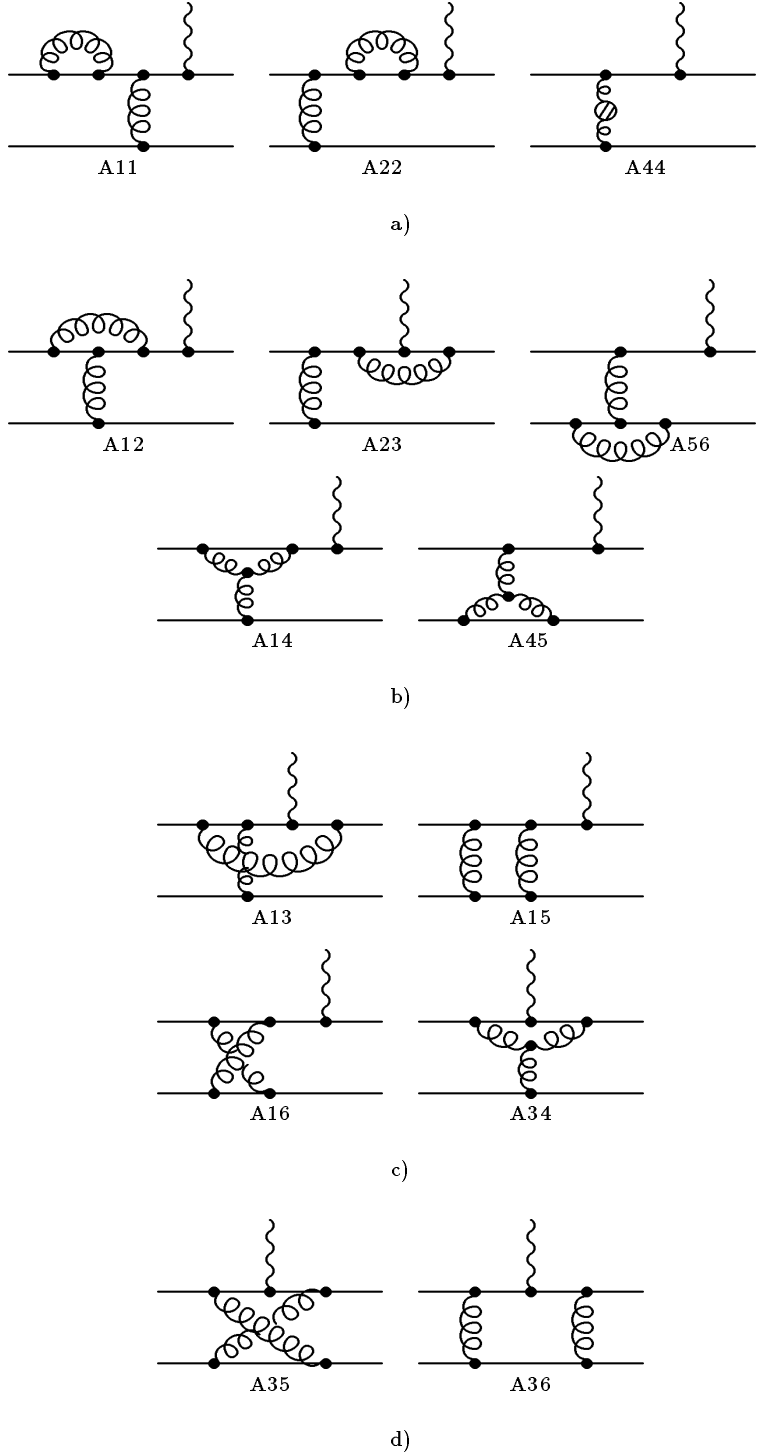,height=20cm,width=10.5cm,silent=}}
 \caption{Distinct one-loop Feynman diagrams contributing to the
    $(q_1 \overline{q}_2) + \gamma^* \rightarrow (q_1 \overline{q}_2)$
    amplitude. The total number of diagrams is 62.}
 \label{f:oldi}
\end{figure}

\widetext
Before the subtraction of divergences has been performed, the pion form factor 
convolution formula reads
\Beq
    F_{\pi}(Q^{2})=\int_{0}^{1} du \int_{0}^{1} dv \;
                    {\sl \Phi}^{*}(v) \;
             \Delta (u,v,\mu_{UV}^2/Q^2,\mu_{IR}^2/Q^2)   \;
                    {\sl \Phi}(u) \, ,
\label{eq:piffcf1}
\Eeq
where $ \Delta (u,v,\mu_{UV}^2/Q^2,\mu_{IR}^2/Q^2) $ denotes
the amplitude for the 
$(q_1 \overline{q}_2) + \gamma^* \rightarrow (q_1 \overline{q}_2)$
quark subprocess. This amplitude has the expansion of the form 
\Beq
   \Delta (u,v,\mu_{UV}^2/Q^2,\mu_{IR}^2/Q^2)=
       \alpha_{S} \, \Delta^{(0)}(u,v,Q^2)
       + \frac{\alpha_S^2}{\pi} 
	  \Delta^{(1)} (u,v,\mu_{UV}^2/Q^2,\mu_{IR}^2/Q^2) + ... \, ,
\label{eq:delta}
\Eeq
where
\Beq
   \Delta^{(0)}(u,v,Q^2)=(1-\epsilon) \, T_{H}^{(0)} (u,v,Q^2) \, .
\label{eq:delta0}
\Eeq

\narrowtext

The contributions
of the individual diagrams to $\Delta^{(1)}(u,v,\mu_{UV}^2/Q^2,
\mu_{IR}^2/Q^2)$, with the overall factor $T_H^{(0)}(u,v,Q^2)$ of 
Eq. \req{eq:TH0} extracted, are listed in Table \ref{t:aresults}.
We have used the following notation:
\Beqa
    \uu&=&1-u  , \; \vv=1-v \nonumber \, ,\\
    \UV&=&\frac{1}{\eUV}-\gamma-\ln \frac{Q^{2}}{\Nuv} 
	= \frac{1}{\Euv} + \ln \frac{\mu_{UV}^2}{Q^2} \, , \nonumber \\
    \IR&=&\frac{1}{\eIR}+\gamma+\ln \frac{Q^{2}}{\Nir} 
	= \frac{1}{\Eir} - \ln \frac{\mu_{IR}^2}{Q^2} \, , \nonumber \\
    \ir&=&\frac{1}{\eIR^{2}}+\frac{1}{\eIR}
          \left( \gamma + \ln \frac{Q^{2}}{\Nir} \right) +
          \frac{1}{2} \left( \gamma^{2}-\frac{\pi^{2}}{6} \right)
             \nonumber \\ & &
          + \gamma \ln \frac{Q^{2}}{\Nir} 
          + \frac{1}{2} \ln^{2} \frac{Q^{2}}{\Nir} \, , \nonumber \\
    \frac{1}{\Euv}&=&\frac{1}{\eUV}-\gamma+\ln (4 \pi) \, , \nonumber \\
    \frac{1}{\Eir}&=&\frac{1}{\eIR}+\gamma-\ln (4 \pi) \, . 
\label{eq:eta}
\Eeqa 
 
\mediumtext

\begin{table}
\caption{Contributions  to $\Delta^{(1)}(u,v,\mu_R^2/Q^2,\mu_F^2/Q^2)$ 
         (defined by \protect\req{eq:delta}) of Feynman diagrams 
         shown in Fig. \protect\ref{f:oldi} .}
\begin{tabular}{cl} 
 A11 & $ \displaystyle -\frac{1}{12}  (\UV+\IR)  $  
			 \\[.4cm]         
 A22 & $ \displaystyle -\frac{1}{6}  (\UV - \ln \vv)  $  
			\\[.4cm]         
 A44 & $ \displaystyle \frac{1}{8}  \left[ \left( 5-\frac{2}{3} n_{f} \right)
              ( \UV - \ln (\uu \vv) ) + \frac{16}{3} -
               \frac{4}{9} n_{f} \right] 
	               $  \\[.4cm] 
 A23 & $ \displaystyle \frac{1}{6} \left( \UV + \frac{1+\vv}{v} \ln \vv \right)
                       $  \\[.4cm]   
 A12 & $ \displaystyle -\frac{1}{48} \left[ \UV + 2 \IR 
                 \left( 1+ \frac{\uu}{u} \ln {\uu} \right)
                 -3-\frac{\uu}{u} \ln {\uu} + \ln {\vv} 
                 + \frac{\uu}{u} \ln^{2} \uu + 
                 2 \frac{\uu}{u} \ln{\uu} \ln{\vv} \right]
                          $  \\[.4cm]   
 A56 & $ \displaystyle -\frac{1}{48} [ \UV + 2 \IR (1- \ln ( \uu \vv))
              -2 \ir 
              -5+ \ln (\uu \vv) - \ln^2 \uu - \ln^2 \vv
              -2 \ln \uu \ln \vv ]
                     $  \\[.4cm]   
 A14 & $  \displaystyle \frac{3}{16} \left[ 3 \UV + 2 \IR 
                \left( 1 + \frac{\ln {\uu}}{u} \right) 
                + 1 + \ln \frac{\uu}{\vv} +\frac{\ln^{2} \uu}{u} 
                + 2 \frac{\ln \uu \ln \vv}{u} \right]
                   $  \\[.4cm]   
 A45 & $ \displaystyle  \frac{3}{16} [ 3 \UV + 4 \IR - 1 + \ln (\uu \vv) ]
		   $  \\[.4cm]   
 A15 & $ \displaystyle  -\frac{1}{3} \frac{\uu}{u}   
              \left[ \IR \ln \uu + \ln \uu +
              \frac{1}{2} \ln^{2} \uu +
              \ln \uu \ln \vv \right]
                     $  \\[.4cm]   
 A16 & $  \displaystyle - \frac{1}{24}  \left[
               \ir + \IR \left( 1 + \frac{\ln \uu}{u} + 2 \ln u 
                            + \ln \vv \right) 
                + \frac{\ln \uu}{u} + 2 \ln u + \ln \vv
                + \frac{1}{2} \frac{\ln^{2} \uu}{u} +
                \ln^{2} u 
                   \right. $  \\ & $ \displaystyle  \left.
		+ \frac{1}{2} \ln^{2} \vv
                + \frac{\ln \uu \ln \vv}{u} 
                + 2 \ln u \ln \vv \right]
                    $  \\[.4cm]   
 A34 & $ \displaystyle \frac{3}{8} \left[ \IR \left( 2 + \frac{\ln \uu}{u} +
                                            \frac{\ln \vv}{v} \right)
              -2 + \ln \uu + \ln \vv 
               +  \frac{\uu +\vv }{u v} \ln \uu \ln \vv 
	       + \frac{1} {2 u} \ln^{2} \uu 
	       + \frac{1} {2 v} \ln^{2} \vv \right]
                  $  \\[.4cm]   
 A13 & $ \displaystyle \frac{1}{24} \left\{ \ir 
            + \IR \left( \frac{\uu}{u} \ln \uu + \ln u + \ln v \right) 
            + 1 
                 \right. $  \\ & $ \displaystyle  \left.
            + \frac{\uu}{2 u} \ln^{2} \uu + \frac{1}{2} \ln^{2} u
            + \frac{1}{2} \ln^{2} v
            + \ln u \ln v + \frac{\uu}{u} \ln \uu \ln \vv 
                  \right. $  \\ & $ \displaystyle  \left.
            - \frac{1}{2 (u-v)^{2}} \left[
               4 \uu^{2} u \mbox{H}(u,\vv) 
            + (4u-5u^{2}+v^{2}) (\ln u + \ln v) 
                  \right. \right. $  \\ & $\displaystyle \left. \left.
            + (6u-5u^{2}-2v+v^{2}) \uu \frac{\ln \uu}{u}
            + (2u-4u^{2}+2v-4uv+5u^{2}v-v^{3}) \frac{\ln \vv}{v}
                  \right] \right\}
		  $  \\[.4cm]   
 A36 & $ \displaystyle -\frac{1}{3} \left[ \IR \left(
             \frac{\uu}{u} \ln{\uu} + \frac{\vv}{v} \ln \vv \right)
              + \frac{\uu}{u} \ln \uu + \frac{\vv}{v} \ln \vv
              + \frac{\uu}{2 u} \ln^{2} \uu
              + \frac{\vv}{2 v} \ln^{2} \vv
              + \frac{\uu+\vv}{u v} \ln \uu \ln \vv \right]
                   $  \\[.4cm]   
 A35 & $ \displaystyle -\frac{1}{24} \left[ 2 \ir
              + \IR \left( 2 + \frac{1+u}{u} \ln \uu
                             + \frac{1+v}{v} \ln \vv \right)
              - 2 (1-u-v) \mbox{H} (u,v)
              + \frac{1+u}{u} \ln \uu
                   \right.  $  \\  & $ \displaystyle  \left.
              + \frac{1+v}{v} \ln \vv
              + \frac{1+u}{2 u} \ln^{2} \uu
              + \frac{1+v}{2 v} \ln^{2} \vv
              + \frac{\uu + \vv}{u v} \ln \uu \ln \vv \right]
		   $  \\  
\end{tabular}
\label{t:aresults}
\end{table}

\widetext

\narrowtext

The function $\mbox{H}(u,v)$ appearing in Table \ref{t:aresults}
is given by the expression
\Beqa
    \mbox{H}(u,v)&=&\frac{1}{1-u-v} \left[
               \Li \left( \frac{\vv}{u} \right)
              + \Li \left( \frac{\uu}{v} \right)
              + \Li \left( \frac{u v}{\uu \vv} \right)
                     \right. \nonumber \\ & & \left.
              - \Li \left( \frac{u}{\vv} \right)
              - \Li \left( \frac{v}{\uu} \right)
              - \Li \left( \frac{\uu \vv}{u v} \right)
                     \right]  \, ,
\label{eq:H}
\Eeqa 
where $\Li (u)$ is the Spence function defined as
\Beq
   \Li (u)=-\int_{0}^{u} \frac{\ln (1-t)}{t} dt  \, .
\Eeq


Adding up the contributions of all diagrams, we find that the NLO 
contribution to the 
$(q_1 \overline{q}_2) + \gamma^* \rightarrow (q_1 \overline{q}_2)$
amplitude, given in (\ref{eq:delta}), can be written in the form 
\Beqa
    \lefteqn{\Delta^{(1)}(u,v,\mu_{UV}^2/Q^2,\mu_{IR}^2/Q^2)}
		   \nonumber \\ & \quad = & 
                 T_H^{(0)}(u,v,Q^2)  \,
	           \left[
	      C_{UV} \frac{1}{\Euv} + C_{IR}(u,v) \frac{1}{\Eir}   
                   \right.  \nonumber \\ & \quad & \left.
	      + \widetilde{f}_{UV}(u,v,\mu_{UV}^2/Q^2) +
            \widetilde{f}_{IR}(u,v,\mu_{IR}^2/Q^2) +
	      \widetilde{f}_{C}(u,v) \right] \, ,
\label{eq:delta1}
\Eeqa
where
\Bml
\label{eq:Cdef}
\Beqa
	C_{UV}&=&\frac{1}{4} \beta_{0} \, , 
	        \\ [0.2cm] 
	C_{IR}(u,v)&=&\frac{2}{3} \left[3+ \ln (\uu \vv) \right] \, , 
\Eeqa
\Eml
and
\Bml
\label{eq:ftdef}
\Beqa
        \widetilde{f}_{UV}(u,v,\mu_{UV}^2/Q^2)&=&C_{UV}
	     \left[  \frac{2}{3} - \ln (\uu \vv) 
	             + \ln \frac{\mu_{UV}^2}{Q^2} \right]
	      \, ,  \label{eq:ftUVdef}\\  [0.2cm] 
	\widetilde{f}_{IR}(u,v,\mu_{IR}^2/Q^2)&=& C_{IR}(u,v) 
             \left[ \frac{1}{2} \ln (\uu \vv) 
	           - \ln \frac{\mu_{IR}^2}{Q^2} \right]
            \, ,  \label{eq:ftIRdef}\\ [0.2cm]  
        \widetilde{f}_{C}(u,v)&=&\frac{1}{12} \left[ -10 
               +20 \ln (\uu \vv) 
               + \ln u \ln v 
               \right. \nonumber \\   & & \left.
               +  \ln \uu \ln \vv
               - \ln u \ln \vv - \ln \uu \ln v
               \right. \nonumber \\  & & \left.
               + (1-u-v) \mbox{H} (u,v) + \mbox{R} (u,v) \right] \, .
             \label{eq:ftCdef}
\Eeqa
\Eml

The function $\mbox{R}(u,v)$ is defined as
\Beqa
   \mbox{R}(u,v)&=&\frac{1}{(u-v)^{2}} \left[
          (2uv-u-v) (\ln u + \ln v)
               \right. \nonumber \\ & & \left.
          + (-2uv^{2}-2v^{2}+10uv-2v-4u^{2}) \frac{\ln \vv }{v}
               \right. \nonumber \\ & & \left.
          + (-2vu^{2}-2u^{2}+10uv-2u-4v^{2}) \frac{\ln \uu }{u}
               \right. \nonumber \\ & & \left.
          - (v \vv^{2}+u \uu^{2}) \mbox{H} (u,\vv) \right]  \, .
\label{eq:R}
\Eeqa 

The contributions of the individual diagrams listed in Table \ref{t:aresults}
are in agreement 
with those obtained in Ref. \cite{FiG81} 
(up to some typographical errors listed 
in \cite{BraT87}).
Correspondence between our results and those of Ref. \cite{FiG81}
is established by multiplying the latter by
$(P+P')^{\mu}/Q^2$.

It is easily seen from Eqs. (\ref{eq:delta1}-\ref{eq:R}) that
in summing up the contributions of all diagrams to
the exclusive 
$(q_1 \overline{q}_2) + \gamma^* \rightarrow (q_1 \overline{q}_2)$
amplitude, the
originally present soft singularities (double $1/\epsilon_{IR}^2$ poles)
cancel out, as required. The final result for the amplitude contains
$1/\epsilon_{UV}$ and only simple $1/\epsilon_{IR}$ poles.

We now proceed to treat these poles.
When doing this, one has to exercise some care, since the subtractions
have to be performed in such a way that the universality of the distribution
amplitude as well as the universality of the coupling constant are
simultaneously preserved.

\widetext

\begin{figure}
  \centerline{\epsfig{file=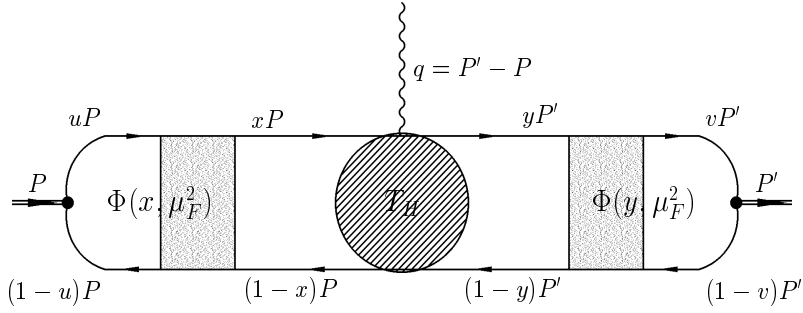,height=4.5cm,width=11cm,silent=}}
 \caption{Pictorial representation of the procedure for absorbing the 
 collinear divergences into the pion distribution amplitude 
 $\Phi(x, \mu_F^2)$.}
 \label{f:piskota}
\end{figure}

The IR poles in \req{eq:delta1}, related to collinear singularities, 
are such that they can be 
absorbed into the pion distribution amplitude.
The universality of the distribution amplitude requires
that the
poles $1/\epsilon_{IR}$ should be absorbed by some universal 
renormalization factor \cite{KhaR85,CuF80}.
The regularization-dependent terms represent ``soft'' effects and
therefore we absorb them into the distribution amplitude along with the 
singularities.
A crucial observation is that the structure of the collinearly
divergent terms in \req{eq:delta1} is such that one can write
\Beqa
   & & V_1(v,y) \otimes T_H^{(0)}(x,y,Q^2) \otimes \delta(x-u) +
    \delta(v-y) \otimes T_H^{(0)}(x,y,Q^2) \otimes V_1(x,u) 
          \nonumber \\   & &\qquad \qquad    
    = C_{IR}(u,v) T_H^{(0)}(u,v,Q^2)  
    \, ,
\label{eq:CIRTH0}
\Eeqa
where the function $V_1(x,u)$ is the one-loop evolution kernel 
for the pion distribution amplitude
(see the corresponding expression in, for example, \cite{FiG81} or 
\cite{BrD86}),
while $\otimes$ denotes the convolution symbol defined as
\Beq
    A(z) \otimes B(z) = \int_0^1 dz \, A(z) \, B(z)
     \: .
\Eeq
Based on Eq. \req{eq:CIRTH0}, the NLO expression for the
$(q_1 \overline{q}_2) + \gamma^* \rightarrow (q_1 \overline{q}_2)$
amplitude,
given by \req{eq:delta}, (\ref{eq:delta1}-\ref{eq:R}),
can be written in a factorized form
\Beqa
      \lefteqn{ \alpha_{S} \, \Delta^{(0)}(u,v,Q^2; \eIR)
       + \frac{\alpha_S^2}{\pi} 
	  \Delta^{(1)} (u,v,\mu_{IR}^2/Q^2; 1/\eIR)}  
           \nonumber \\ & \quad  = & 
     \left[ \delta(v-y) + \frac{\alpha_S}{\pi} V_1(v,y) 
         \left( \frac{1}{\Eir} + \ln \frac{\nIR}{\mu_{IR}^2} \right) \right]
	   \nonumber \\ 
	 & \qquad {} & \otimes 
     \left\{ \alpha_S T_{H}^{(0)}(x,y,Q^2) \left[ (1-\epsilon) + 
        \frac{\alpha_S}{\pi} \widehat{T}_{H}^{(1)}(x,y,\nIR/Q^2;\eIR) 
	                                   \right] \right\} 
	   \nonumber \\ 
	 & \qquad {} & \otimes 
     \left[ \delta(x-u) + \frac{\alpha_S}{\pi} V_1(x,u) 
         \left( \frac{1}{\Eir} + \ln \frac{\nIR}{\mu_{IR}^2} \right) \right]
	  \, .
\label{eq:IRsub}
\Eeqa
Formula \req{eq:IRsub} shows the essence of the procedure for the
separation of collinear divergences, 
which is schematically represented in Fig. \ref{f:piskota}. 
The procedure 
consists in summing the effects of collinear divergences contained in the 
shaded rectangular region in Fig. \ref{f:piskota}, as a result of which 
the outside quark (antiquark) with momentum $u P$ ($\uu P$)
ends up on the inside quark (antiquark) of 
momentum $x P$ ($\x P$).

As for the UV poles, we renormalize them using 
the modified minimal-subtraction ($\overline{\mbox{MS}}$)
scheme. This is carried out by stating that $\alpha_{S}$
appearing in \req{eq:delta} is the bare unrenormalized coupling
related to the renormalized physical coupling 
$\alpha_{\overline{MS}}(\nUV)$ by
\Beq
  \alpha_{S}=\alpha_{\overline{MS}}(\nUV) 
      \left[1-\frac{\alpha_{\overline{MS}}(\nUV)}{\pi}
	      \frac{\beta_{0}}{4} 
	    \left( \frac{1}{\Euv} - \ln\frac{\nUV}{\mu_{UV}^2} \right) 
      \right] \, .
\label{eq:alphaMS}
\Eeq

\widetext

It is important to note that the $\epsilon \rightarrow 0$ limit
can be taken only after the separation of collinear divergences and
UV renormalization have been  performed 
(taking the $\epsilon \rightarrow 0$ limit in $\Delta^{(0)}$
 before the subtraction of IR and UV singularities is equivalent
 to choosing a factorization scheme that does not respect the
 universality of the distribution amplitude (see \cite{KhaR85}) and
 the renormalization scheme in which the running coupling constant 
 is not universal).
As a result, we obtain the following expression for the NLO
hard-scattering amplitude for the pion form factor:
\Beqa
  \lefteqn{T_{H}(x,y,Q^2,\nUV,\nIR)=
   \alpha_{\overline{MS}}(\nUV) \, T_{H}^{(0)}(x,y,Q^2)}
            \nonumber \\ & & \qquad \qquad \times \;
   \left[ 1 + \frac{\alpha_{\overline{MS}}(\nUV)}{\pi} \, 
	       T_{H}^{(1)}(x,y,\nUV/Q^2,\nIR/Q^2) \right] \, ,
\label{eq:THalphaMS}
\Eeqa
where
\Beqa
     \lefteqn{T_{H}^{(1)}(x,y,\nUV/Q^2,\nIR/Q^2) }
                    \nonumber \\ & = &
	 f_{UV}(x,y,\nUV/Q^2)+f_{IR}(x,y,\nIR/Q^2)+f_{C}(x,y)
	 \, ,
\label{eq:TH1}
\Eeqa
and
\Bml
\label{eq:fdef}
\Beqa
        f_{UV}(x,y,\nUV/Q^2) & = &
	     \widetilde{f}_{UV}(x,y,\nUV/Q^2) + C_{UV} \nonumber \\
	     &=& \frac{1}{4} \beta_0
	     \left[  \frac{5}{3} - \ln (\x \y) + \ln \frac{\nUV}{Q^2} \right]
	      \, ,  \label{eq:fUVdef}\\  [0.2cm] 
	f_{IR}(x,y,\nIR/Q^2) & = &
	     \widetilde{f}_{IR}(x,y,\nIR/Q^2) \nonumber \\
	     &=& \frac{2}{3} \left[3+ \ln (\x \y) \right] 
             \left[ \frac{1}{2} \ln (\x \y) - \ln \frac{\nIR}{Q^2} \right]
            \, ,  \label{eq:fIRdef}\\ [0.2cm]  
        f_{C}(x,y) & = &
	     \widetilde{f}_{C}(x,y) - C_{IR}(x,y) \nonumber \\
	     &=&\frac{1}{12} \left[ -34 
               +12 \ln (\x \y) 
               + \ln x \ln y 
               \right. \nonumber \\   & & \left.
               +  \ln \x \ln \y
               - \ln x \ln \y - \ln \x \ln y
               \right. \nonumber \\  & & \left.
               + (1-x-y) \mbox{H} (x,y) + \mbox{R} (x,y) \right] \, .
             \label{eq:fCdef}
\Eeqa
\Eml
\narrowtext

A few comments on the previously performed calculations are in order. 
Our final expression for the hard-scattering amplitude $T_H$ is in agreement
with Ref. \cite{Sa82}. The calculation of \cite{FiG81} contains
a few typographical errors in the
diagram-by-diagram listing as well as in the final expression
(which are all listed in \cite{BraT87}). Apart from this, the results from Ref. \cite{FiG81} agree with ours. 
In Refs. \cite{DiR81,KhaR85} the contribution of the diagrams with propagator
corrections on external quark lines were not taken into account.
Also, in Ref. \cite{DiR81} the subtraction of
collinear singularities was performed in a way that is not consistent
with the universality of the distribution amplitude (see the discussion in
Ref. \cite{KhaR85}).
Finally, a thorough analysis of the results
of Refs. \cite{FiG81,DiR81,Sa82} was performed in Ref. \cite{BraT87},
but in obtaining the final
expression for the hard-scattering amplitude $T_H$, 
collinear and UV divergences were subtracted
in a way that is not consistent with the universality of both the 
distribution amplitude and the running coupling constant.

\section{Evolutional corrections to the pion distribution
amplitude}
\label{sec:DAs}

The pion distribution amplitude ${\sl \Phi}(x,\nIR)$,
controlling exclusive pion processes at large momentum transfer,
is the basic
valence wave function of the pion.
Its form is not yet accurately known.
It has been shown, however, that the leptonic decay
$\pi^+ \rightarrow \mu^+ \nu_{\mu}$ imposes on ${\sl \Phi}(x,\nIR)$ a constraint 
of the form
\Beq
     \int_0^1 dx \,  {\sl \Phi}(x,\nIR) = 
            \frac{f_{\pi}}{2 \sqrt{2 n_C}} \, .
\label{eq:Phinorm}
\Eeq
Given the form of ${\sl \Phi}(x,\nIR)$, this relation normalizes it for any
$\nIR$.
In \req{eq:Phinorm}, $f_{\pi}=0.131$ GeV is the pion decay constant
and $n_C$($=3$) is the number of QCD colors.

Instead of using ${\sl \Phi}(x, \nIR)$ satisfying \req{eq:Phinorm},
one usually introduces the distribution amplitude
$\phi(x, \nIR)$ normalized to unity,
\Beq
     \int_0^1 dx \,  \phi(x,\nIR) = 1 \, , 
\label{eq:phinorm}
\Eeq
and related to ${\sl \Phi}(x, \nIR)$ by
\Beq
     {\sl \Phi}(x, \nIR) = \frac{f_{\pi}}{2 \sqrt{2 n_C}}
                   \phi(x,\nIR) \, .
\Eeq
 
Although intrinsically nonperturbative, the pion distribution amplitude
$\phi(x,\nIR)$ satisfies an evolution equation of the
form
\Beq
  \nIR \frac{\partial}{\partial \nIR} \phi(x,\nIR)   = 
  \int_0^1 du \; V(x,u,\alpha_S(\nIR)) \; \phi(u,\nIR)
         \, ,
\label{eq:eveq}
\Eeq
in which the evolution kernel is calculable in
perturbation theory:
\Beqa
  \lefteqn{V(x,u,\alpha_S(\nIR))} 
           \nonumber \\ & & \quad
        =  \frac{\alpha_S(\nIR)}{\pi} \, V_1(x,u) +
                \left( \frac{\alpha_S(\nIR)}{\pi} \right)^2 V_2(x,u) +
                 \cdots 
           \nonumber \\ & & 
\label{eq:kernel}
\Eeqa
and has been computed in the one- and two-loop approximations
using dimensional regularization and the $\overline{MS}$ scheme.

If the distribution amplitude $\phi(x,\nO)$ 
can be calculated at an initial momentum scale $\nO$ using QCD sum rules
\cite{ChZ84}  or lattice gauge theory \cite{MarS87etc},
then the differential-integral evolution equation \req{eq:eveq} 
can be integrated using the moment method to give $\phi(x,\nIR)$
at any momentum scale $\nIR > \nO$. 

Because of the complicated structure of the two-loop contribution
to the evolutional kernel $V_2(x,y)$, only the first few moments of the
evolutional kernel have been computed numerically.

Recently, based on the conformal spin expansion and the conformal
consistency relation, the analytical result for the evolution
of the flavor-nonsinglet meson distribution amplitude has been
determined \cite{Mu94etc}.

\widetext

To the NLO approximation, this, rather complicated solution has the form
\Beq
   \phi(x,\nIR)=\phi^{LO}(x,\nIR) + \frac{\alpha_S(\nIR)}{\pi} \,
               \phi^{NLO}(x,\nIR) \, ,
\label{eq:evDA}
\Eeq
where
\Beqa
   \phi^{LO}(x,\nIR)  & = &x \, (1-x)
         \sum_{n=0}^{\infty} {}' \:
         C_n^{3/2}(2 x -1) \; 
         \frac{4 (2n+3)}{(n+1) (n+2)} \, 
             \nonumber \\ & & \times \, 
         \left( \frac{\alpha_S(\nO)}{\alpha_S(\nIR)} 
         \right)^{\gamma_n^{(0)}/\beta_0} \,
         \int_0^1 dy \, C_n^{3/2}(2 y -1) \, \phi(y,\nO) \, ,
\label{eq:phiLO}
\Eeqa
and
\Beqa
  \phi^{NLO}(x,\nIR) & = & x \, (1-x) \, \sum_{n=0}^{\infty} {}' \:
         \left( \frac{\alpha_S(\nO)}{\alpha_S(\nIR)} 
         \right)^{\gamma_n^{(0)}/\beta_0} \,
         \int_0^1 dy \, C_n^{3/2}(2 y -1) \, \phi(y,\nO)
             \nonumber \\ & &
       \left[ C_n^{3/2}(2 x -1) \frac{ (2n+3)}{(n+1) (n+2)}
              \left(1-\frac{\alpha_S(\nO)}{\alpha_S(\nIR)} \right) \, 
              \left( \frac{\gamma_n^{(1)}}{2 \beta_0}
                 + \frac{\beta_1}{\beta_0^2} \gamma_n^{(0)} \right)
             \right. \nonumber \\ & & \left.
          +  \sum_{k=n+2}^{\infty} {}' \:
              C_k^{3/2}(2 x -1) \frac{2 (2k+3)}{(k+1) (k+2)}
          S_{kn}(\nIR) \, C_{kn}^{(1)}
                  \right]
               \, ,
\label{eq:phiNLO}
\Eeqa
with abbreviations
\Beqa
    S_{kn}(\nIR) & = &
 \frac{\gamma_k^{(0)}-\gamma_n^{(0)}}{\gamma_k^{(0)}-\gamma_n^{(0)}+\beta_0}
 \left[ 1 - \left( \frac{\alpha_S(\nO)}{\alpha_S(\nIR)}
          \right)^{1+(\gamma_k^{(0)}-\gamma_n^{(0)})/\beta_0} 
 \right] \, ,
               \\
    C_{kn}^{(1)} & = & (2 n + 3)
 \left[ \frac{\gamma_n^{(0)}-\beta_0+4 C_F A_{kn}}{(k-n)(k+n+3)}
    + \frac{ 2 C_F (A_{kn} - \psi (k+2) + \psi(1) )}{(n+1)(n+2)} \right] \, ,
              \nonumber  \\ & & \label{eq:Ckn1} \\
    A_{kn} &=& \psi(\frac{k+n+4}{2})-\psi(\frac{k-n}{2})+
              2 \psi(k-n) - \psi(k+2)- \psi(1) \, ,
                      \label{eq:Akn}
\Eeqa
and $\sum {}'$ denoting the sum running only over even $n$,
while $C_n^{3/2}(z)$ are Gegenbauer polynomials of order $3/2$
and $C_F=4/3$.
\narrowtext
In Eqs. (\ref{eq:phiLO}--\ref{eq:Ckn1}), $\gamma_n^{(0)}$ are the usual
anomalous dimensions:
\Beq
  \gamma_n^{(0)}  =  C_F \left[
            3 + \frac{2}{(n+1) (n+2)}
           - 4 \sum_{i=1}^{n+1} \frac{1}{i} \right] \, ,
\Eeq
while $\beta_0$ and $\beta_1$ are the first two terms
in the expansion of the QCD $\beta$--function, with $\beta_0$ given 
by \req{eq:beta0}, and 
\Beq
    \beta_1=102-\frac{38}{3} n_f \, .
\Eeq
The function $\psi(z)$ appearing in \req{eq:Ckn1} and
\req{eq:Akn} is defined as
\Beq
    \psi(z)= \frac{d}{dz} (\ln \Gamma(z)).
\Eeq
For $n_f=3$ and for the values of $n$ we are interested in,
we have taken the values of the anomalous dimension $\gamma_n^{(1)}$
from \cite{FlR77etc,GoAL79}:
\Beq
    \gamma_0^{(1)}=0 \, , \qquad 
    \gamma_2^{(1)}=111.03 \, , \qquad 
     \gamma_4^{(1)}=150.28 \, .
\Eeq

\widetext

To proceed, we expand the distribution amplitude
$\phi(x,\mu_0^2)$ in terms of the
Gegenbauer polynomials $C_n^{3/2}(2 x -1)$ 
(the eigenfunctions of the LO evolution equation)
\Beq
     \phi(x)=\phi(x,\mu_0^2)= 6 x (1-x) \,
	  \sum_{n=0}^{\infty} {}' B_n \:  C_n^{3/2}(2 x - 1) 
	   \, .
\label{eq:phiexp}
\Eeq
Owing to the orthogonality of the Gegenbauer polynomials,
the expansion coefficients $B_n$ are given by
\Beq
B_n=\frac{1}{6} \, \frac{4 (2 n + 3)}{(n+1)(n+2)} \,
       \int_0^1 dx \, C_n^{3/2}(2 x-1) \, \phi(x) \,
\label{eq:Bn}
\Eeq
and the normalization condition \req{eq:phinorm} 
then implies
$B_0=1$.
Substituting \req{eq:phiexp}
into \req{eq:phiLO}
gives
\Beq
     \phi^{LO}(x,\nIR)= 6 x (1-x) \,
	  \sum_{n=0}^{\infty} {}' B_n^{LO}(\nIR) \:  C_n^{3/2}(2 x - 1) 
	  \, ,
\label{eq:phiLOexp}
\Eeq
where 
\Beq
        B_n^{LO}(\nIR)  =  B_n 
             \left( \frac{\alpha_S(\mu_0^2)}{\alpha_S(\nIR)} 
             \right)^{\gamma_n^{(0)}/\beta_0} \, ,
\label{eq:BnLO}
\Eeq
so that, obviously, $B_0^{LO}=1$.
Next, by taking \req{eq:phiexp} into account
\req{eq:phiNLO} becomes
\Beq
     \phi^{NLO}(x,\nIR) =  6 x (1-x) \, 
	  \sum_{k=2}^{\infty} {}' B_k^{NLO}(\nIR) \; C_k^{3/2}(2 x - 1) 
	  \, ,
\label{eq:phiNLOexp}
\Eeq
where
\Beq
        B_k^{NLO}(\nIR)  =   
             B_k^{LO}(\nIR) P_{k}(\nIR)
	     + \sum_{n=0}^{k-2} {}' 
             B_n^{LO}(\nIR) Q_{kn}(\nIR) 
	     \, ,
\label{eq:BnNLO}
\Eeq
and
\Beqa
  P_{k}(\nIR) & = & \frac{1}{4} \, 
              \left( \frac{\gamma_k^{(1)}}{2 \beta_0}
                 + \frac{\beta_1}{\beta_0^2} \gamma_k^{(0)} \right) \,
              \left(1-\frac{\alpha_S(\nO)}{\alpha_S(\nIR)} \right) 
	       \, ,
\label{eq:I}
	       \\  
  Q_{kn}(\nIR) & = & \frac{(2k+3)}{(k+1)(k+2)} \,
                \frac{(n+2)(n+1)}{2 (2n+3)} \,  
		C_{kn}^{(1)} \, S_{kn}(\nIR) \, .
\label{eq:II}
\Eeqa

\begin{figure}
   \centerline{\epsfig{file=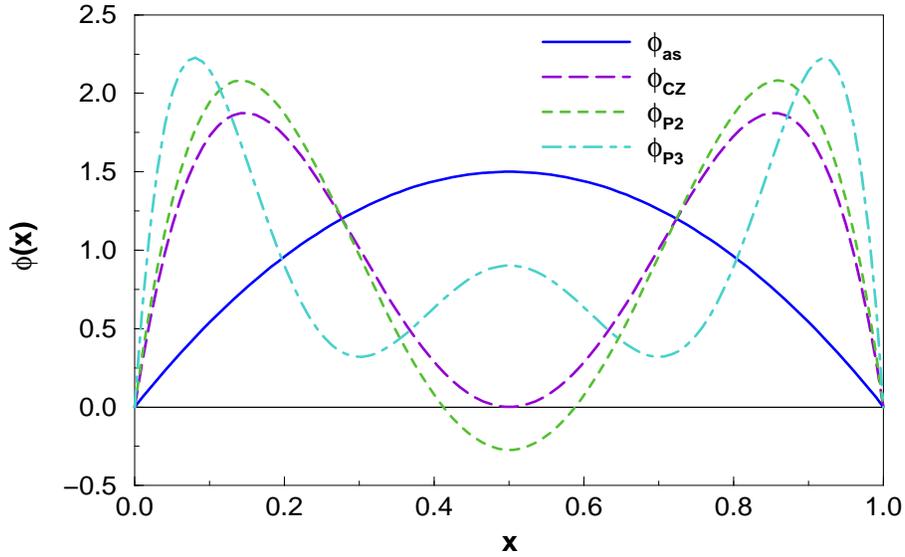,height=8cm,width=12cm,silent=}}
\caption{The four candidate pion distribution amplitudes 
         defined by Eqs. \protect\req{eq:phix},
         chosen as nonperturbative inputs at the reference scale
         $\mu_0^2=(0.5$GeV$)^2$.}
\label{f:xDA}
\end{figure}

For the purpose of our calculation,
we use the following four 
candidate distribution amplitudes
(shown in Fig. \ref{f:xDA})
as nonperturbative inputs at the reference momentum
scale $\nO=(0.5$ GeV$)^2$: 
\Bml
\label{eq:phix}
\Beqa
   \phi_{as}(x) & \equiv & \phi_{as}(x,\nO) 
               \, = \, 6 x (1 - x) \, ,
                \label{eq:phixAS}      \\
   \phi_{CZ}(x) & \equiv & \phi_{CZ}(x,\nO)
            \, = \, 6 x (1 - x) \left[ 5 (2 x-1)^2 \right] \, ,
                 \label{eq:phixCZ}     \\
   \phi_{P2}(x) & \equiv & \phi_{P2}(x,\nO) 
             =  6 x (1 - x)
              \left[ -0.1821 + 5.91 (2 x-1)^2 \right] \, ,
                 \label{eq:phixP2}     \\
   \phi_{P3}(x) & \equiv & \phi_{P3}(x,\nO) 
             =  6 x (1 - x)
              \left[ 0.6016 -4.659 (2 x-1)^2 + 15.52 (2 x-1)^4 \right]
                      \, .
                   \label{eq:phixP3}
\Eeqa
\Eml

\narrowtext

Here $\phi_{as}(x)$ is the asymptotic
distribution amplitude  and represents the solution of the
evolution equation \req{eq:eveq} for $\nIR \rightarrow \infty$.
The double-hump-shaped distribution amplitudes
$\phi_{CZ}(x)$ and $\phi_{P2}(x)$ and the three-hump-shaped
distribution amplitude $\phi_{P3}(x)$ have been obtained using
the method of QCD sum rules 
\cite{ChZ84,FaH91}. 
As Fig. \ref{f:xDA} shows, these distribution amplitudes,
unlike $\phi_{as}(x)$, are strongly end-point concentrated.
In the limit $\nIR \rightarrow \infty$, they
reduce to the asymptotic form $\phi_{as}(x)$.

\widetext

The pion candidate distribution amplitudes given by
\req{eq:phix} are of the general form
\Beq
     \phi(x)=\phi(x,\mu_0^2)=\phi_{as}(x) \,
	   [ 1 + B_2 \, C_2^{3/2}(2 x -1) + B_4 \, C_4^{3/2}(2 x -1) ]
	   \, ,
\label{eq:phi4exp}
\Eeq
with the corresponding coefficients
\Bml
\label{eq:Bcoeff}
\Beqa
   \phi_{as}(x): &  & \qquad B_n=0 \:, n \geq 2
                \label{eq:asBcoeff} \\
   \phi_{CZ}(x): &  & \qquad B_n=0 \:, n > 2 \qquad B_2=2/3 
                 \label{eq:czBcoeff} \\
   \phi_{P2}(x): &  & \qquad B_n=0 \:, n > 2 \qquad B_2=0.788 
                \label{eq:p2Bcoeff} \\
   \phi_{P3}(x): &  & \qquad B_n=0 \:, n > 4 
                     \qquad B_2=0.7582 \qquad B_4=0.3941
                 \label{eq:p3Bcoeff} \, ,
\Eeqa
\Eml
with $B_0=1$.
Now, according to \req{eq:phiLOexp} and \req{eq:phiNLOexp}, 
the LO and NLO parts of the distribution amplitude \req{eq:phi4exp}
read
\Beqa
     \phi^{LO}(x,\nIR) & = & \phi_{as}(x)
	   \left[ 1 + B_2^{LO}(\nIR) \, C_2^{3/2}(2 x -1) 
	       + B_4^{LO}(\nIR) \, C_4^{3/2}(2 x -1) \right]
	       \, ,
\label{eq:phiLO4exp} \\
     \phi^{NLO}(x,\nIR) & = & \phi_{as}(x)
	   \left[ B_2^{NLO}(\nIR) \, C_2^{3/2}(2 x -1) 
	       + B_4^{NLO}(\nIR) \, C_4^{3/2}(2 x -1) 
	       \right. \nonumber \\ & & \left.
	     + \sum_{k=6}^{\infty}{}' B_k^{NLO}(\nIR) \, C_k^{3/2}(2 x -1)
	       \right]
	        \, ,
\label{eq:phiNLO4exp} 
\Eeqa
where 
\Beqa
        B_2^{LO}(\nIR)  =  B_2 
             \left( \frac{\alpha_S(\mu_0^2)}{\alpha_S(\nIR)} 
             \right)^{-50/81} \, ,
	        & \quad &
        B_4^{LO}(\nIR)  =  B_4 
             \left( \frac{\alpha_S(\mu_0^2)}{\alpha_S(\nIR)} 
             \right)^{-364/405}
	     \, ,
\label{eq:BnLO4}
\Eeqa
and
\Beqa
        B_2^{NLO}(\nIR) & = &  
	     B_2^{LO}(\nIR) \, P_{2}(\nIR)
	     + Q_{20}(\nIR) 
	       \, ,
	     \nonumber \\
        B_4^{NLO}(\nIR) & = & 
	     B_4^{LO}(\nIR) \, P_{4}(\nIR)
	     + Q_{40}(\nIR) + B_2^{LO}(\nIR) \, Q_{42}(\nIR)
	       \, ,
	     \nonumber \\
        B_k^{NLO}(\nIR) & = & 
	     Q_{k0}(\nIR) + B_2^{LO}(\nIR) \, Q_{k2}(\nIR)
             + B_4^{LO}(\nIR) \, Q_{k4}(\nIR) \:, k \geq 6
	     \, .
\label{eq:BnNLO4}
\Eeqa
Note that $\phi^{NLO}(x,\nIR)$ represents the infinite sum of
Gegenbauer polynomials even though the $\phi(x)$ distribution
is described by a finite number of terms.
Since $B_k^{NLO}$ decreases with $k$,
for the purpose of numerical calculation one can approximate
$\phi^{NLO}(x,\nIR)$ by neglecting higher-order Gegenbauer polynomials
($k > 100$).

\narrowtext

\begin{figure}
   \epsfig{file=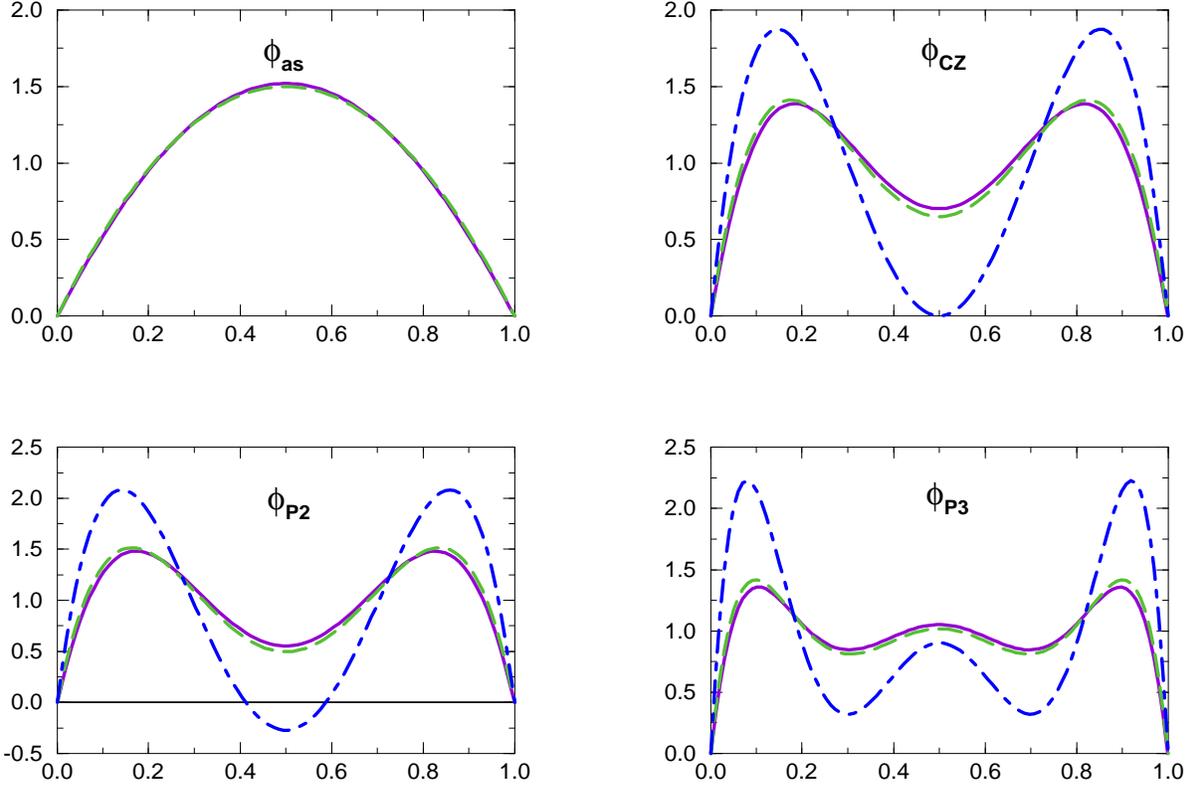,height=11.5cm,width=16.5cm,silent=}
\caption{Evolution of the four candidate  pion distribution
         amplitudes:
         $\phi_{as}(x,\nIR)$,
         $\phi_{CZ}(x,\nIR)$, $\phi_{P2}(x,\nIR)$, 
         and $\phi_{P3}(x,\nIR)$,
         with the running coupling constant $\alpha_S(\nIR)$ and
         three active flavors.
          The dash-dotted curves correspond to the distribution
          amplitudes given by \protect\req{eq:phix}
          and taken as nonperturbative input at the reference momentum scale
          $\nO=(0.5$ GeV$)^2$.
          The dashed curves correspond to the distribution amplitudes
          at the momentum scale $\nIR=(2$ GeV$)^2$ with the LO
          evolutional corrections included according to 
         \protect\req{eq:phiLO4exp} .
          The solid curves correspond to the distribution amplitudes
          at the momentum scale $\nIR=(2$ GeV$)^2$ with the NLO 
          evolutional corrections taken into account
          according to Eqs.
         \protect\req{eq:evDA} and
	 (\protect\ref{eq:phiLO4exp}--\protect\ref{eq:phiNLO4exp}).}
\label{f:DA}
\end{figure}

A summary of our results for the four candidate distribution amplitudes 
with the NLO evolutional corrections included
is shown in Fig. \ref{f:DA}.
The dash-dotted curves correspond to the distribution amplitudes at the
reference point $\nIR=\mu_0^2=(0.5$ GeV$)^2$.
The dashed and solid curves represent the distribution
amplitudes evoluted to $\nIR=(2$ GeV$)^2$, with the difference 
that the former includes only LO evolutional corrections, whereas
the latter includes NLO evolutional corrections.
As it is obvious from \req{eq:asBcoeff} and 
(\ref{eq:phiLO4exp} -- \ref{eq:phiNLO4exp}), 
although $\phi_{as}$ shows no LO evolution, there are tiny
effects of the NLO evolution.
For other distributions, the LO evolution is significant, and even the
NLO evolution is nonnegligible.
Any distribution amplitude evolves asymptotically 
(i.e., for $\nIR \rightarrow \infty$) into $\phi_{as}(x)$, but 
the higher $\nIR$ is, the ``slower'' this approach becomes.

\widetext

\narrowtext

\section{Choosing the factorization and the renormalization
            scales}
\label{sec:scales}

\widetext

In this section we discuss various possibilities
of choosing the renormalization scale
$\mu_{R}$ and the factorization scale $\mu_{F}$
appropriate for the process under consideration.

Before starting the discussion, let us now, 
with the help of Eqs. \req{eq:piffcf}, \req{eq:TH}, and \req{eq:evDA},
write down the complete leading-twist NLO QCD expression
for the pion electromagnetic form factor
with the $\mu_{R}$ and $\mu_{F}$ dependence
of all the terms explicitly indicated.

Generally, for the NLO form factor we can write
\Beq
  F_{\pi}(Q^2,\nUV,\nIR) = F_{\pi}^{(0)}(Q^2,\nUV,\nIR)
          + F_{\pi}^{(1)}(Q^2,\nUV,\nIR) \, .
\label{eq:Fpi}
\Eeq
The first term in \req{eq:Fpi} is the LO contribution 
and is given by 
\Beq
 F_{\pi}^{(0)}(Q^2,\nUV,\nIR) =
       \int_0^1 dx \, \int_0^1 dy  \quad \alpha_S(\nUV) \: 
        \Phi^{LO*}(y,\nIR) \,
              T_H^{(0)}(x,y,Q^2) \,
              \Phi^{LO}(x,\nIR)  \, .
\label{eq:Fpi1}
\Eeq
The second term in \req{eq:Fpi} is the NLO contribution 
and can be written as
\Beq
  F_{\pi}^{(1)}(Q^2,\nUV,\nIR) = F_{\pi}^{(1a)}(Q^2,\nUV,\nIR)
          + F_{\pi}^{(1b)}(Q^2,\nUV,\nIR) \, ,
\label{eq:Fpi2}
\Eeq
where
\Beqa
 \lefteqn{F_{\pi}^{(1a)}(Q^2,\nUV,\nIR)  = 
      \, \int_0^1 dx \, \int_0^1 dy \: \frac{\alpha_S^2(\nUV)}{\pi} \: 
        } \nonumber \\ & & \qquad \qquad
      \times \,
       \Phi^{LO*}(y,\nIR) \,
             T_H^{(0)}(x,y,Q^2) \, T_H^{(1)}(x,y,\nUV/Q^2,\nIR/Q^2) \,
              \Phi^{LO}(x,\nIR)  \, ,
\label{eq:Fpi2a}
\Eeqa
is the contribution coming from the NLO 
correction to the hard-scattering amplitude, whereas
\Beqa
 \lefteqn{F_{\pi}^{(1b)}(Q^2,\nUV,\nIR)  =  
   \int_0^1 dx \, \int_0^1 dy \quad\frac{\alpha_S(\nUV) \alpha_S(\nIR)}{\pi} 
        } \nonumber \\ & & \quad 
    \times \, 
    \left[\Phi^{NLO*}(y,\nIR) \,
           T_H^{(0)}(x,y,Q^2) \,
           \Phi^{LO}(x,\nIR) 
          + \,   \Phi^{LO*}(y,\nIR) \,
           T_H^{(0)}(x,y,Q^2) \,
           \Phi^{NLO}(x,\nIR) \right] 
	   \nonumber \\ & &
\label{eq:Fpi2b}
\Eeqa
is the contribution arising from the inclusion of the  
NLO evolution of the distribution amplitude.
Now, if \req{eq:TH0} is taken into account, the
expression for the LO contribution of Eq. \req{eq:Fpi1}
can be written in the form
\Beq
    F_{\pi}^{(0)}(Q^2,\nUV,\nIR)=
      \frac{8}{9} \, \pi \, \frac{ f_{\pi}^2}{Q^2}  \, 
      \int_0^1 dx \, \int_0^1 dy \quad \alpha_S(\nUV)\:
      \frac{\phi^{LO}(x,\nIR)}{\x} \: 
      \frac{\phi^{LO*}(y,\nIR)}{\y} 
             \, .
\label{eq:Q2Fpi1old}
\Eeq

Next, before going on to perform the remaining $x$ and $y$ integrations,
we have to choose the renormalization scale $\mu_{R}$ 
and the factorization scale $\mu_{F}$.
In doing this, however, there is considerable freedom involved.

\narrowtext

If calculated to all orders in perturbation theory,
the physical pion form factor $F_{\pi}(Q^2)$,
represented at the sufficiently high $Q^2$ by the factorization formula 
\req{eq:piffcf}, would be independent of the 
renormalization and factorization scales, $\mu_R$ and $\mu_F$, 
and so they are arbitrary parameters.
Truncation of the perturbative expansion of $F_{\pi}(Q^2)$
at any finite order causes a residual dependence on 
these scales. 
Although the best choice for these scales remains an open question
(the scale ambiguity problem),
one would like to choose them in such a way that they are of 
order of some physical scale in the problem, and, at the same time,
to reduce the size of higher-order corrections as much as possible.
Choosing any specific value for these scales leads to a
theoretical uncertainty of the perturbative result.

In our calculation we approximate $F_{\pi}(Q^2)$ 
only by  two terms of the perturbative series and
hope that we can minimize higher-order corrections
by  a suitable choice of $\mu_{R}$ and $\mu_{F}$, 
so that the LO term $F_{\pi}^{(0)}(Q^2,\nUV,\nIR)$
gives a good approximation 
to the complete sum $F_{\pi}(Q^2)$.

It should be noted that in $F_{\pi}^{(0)}(Q^2,\nUV,\nIR)$, 
given by \req{eq:Fpi1},
$\mu_{R}$ appears only through $\alpha_S(\nUV)$,
whereas $\mu_{F}$ enters into the distribution amplitude $\phi(x, \nIR)$.
In $F_{\pi}^{(1)}(Q^2,\nUV,\nIR)$, given by (\ref{eq:Fpi2}--\ref{eq:Fpi2b}),
a logarithmic dependence on the scales $\mu_{R}$ and $\mu_{F}$ 
appears also through $T_{H}^{(1)}(x,y,\nUV/Q^2,\nIR/Q^2)$.
As seen from \req{eq:fdef} and \req{eq:TH1},
this dependence is contained in the terms
\Beqa
      f_{UV}(x,y,Q^2/\nUV)&=&\frac{1}{4} \beta_{0}
           \left[  \frac{5}{3} - \ln (\x \y) + \ln \frac{\nUV}{Q^2} \right]
               \, , \label{eq:fUV} \\
      f_{IR}(x,y,Q^2/\nIR)&=&\frac{2}{3} [3+ \ln (\x \y)]
             \left[ \frac{1}{2} \ln (\x \y) - \ln \frac{\nIR}{Q^2} \right]
              \, .  \label{eq:fIR} 
\Eeqa
Being independent of each other, the scales $\mu_R$
and $\mu_F$ can be expressed in terms of $Q^2$ as
\Beqa
      \nUV&=&a(x,y) \, Q^2 \, ,    \\
      \nIR&=&b(x,y) \, Q^2 \, ,
\Eeqa
where $a(x,y)$ and $b(x,y)$ are some linear functions of the dimensionless
variables $x$ and $y$ (quark longitudinal momentum fractions).

We next discuss various possibilities of choosing the scales $\mu_R$ and
$\mu_F$ separately.
The simplest and widely used choice for the scale $\mu_R$
is
\Beq
      \nUV=Q^2 \, ,
\label{eq:nRQ}
\Eeq
the justification for the use of which is mainly pragmatic.

Physically, however, the more appropriate choice for $\nUV$ 
would be that one corresponding to the characteristic
virtualities of the particles in the parton subprocess,
which is considerably lower  than the overall momentum transfer
$Q^2$ (i.e., virtuality of the probing photon).

It follows from Figs. \ref{f:piff} and \ref{f:loan}
that the virtualities of the gluon line (line 4) and
the internal quark line (line 2) of diagram A
in Fig. \ref{f:loan} are given by
$\x \y Q^2$ and $\y Q^2$, respectively.
Now, if instead of using \req{eq:nRQ} we choose
$\mu_R$ to be equal to the gluon virtuality,
i.e.,
\Beq
  \nUV=\x \y Q^2 \, ,
\label{eq:nRg}
\Eeq
then the logarithmic terms in \req{eq:fUV}
vanish.

As it is well known, unlike in an Abelian theory (e.g., QED),
where the effective coupling is entirely renormalized
by the corrections of the vector particle propagator,
in QCD the coupling is renormalized not only by the gluon propagator,
but also by the quark-gluon vertex and quark-propagator
corrections.
It is thus possible to choose $\nUV$ as the geometrical mean
of the gluon and quark virtualities \cite{DiR81}:
\Beq
   \nUV=\sqrt{(\x \y Q^2) (\y Q^2)} \, .
\label{eq:nRgq}
\Eeq

Alternatively, we can make a choice 
\Beq
  \nUV=e^{-5/3} \x \y Q^2 \, ,
\label{eq:nRge}
\Eeq
as a result of which the function $f_{UV}$, given by \req{eq:fUV},
vanishes identically. In this case, 
$T_{H}^{(1)}(x,y,\nUV/Q^2,\nIR/Q^2)$ defined by 
\req{eq:TH1} becomes $n_f$ independent.
This is an example of choosing the renormalization scale
according to the Brodsky-Lepage-Mackenzie (BLM) procedure \cite{BrL83}.
In this procedure, the renormalization scale $\nUV$ best suited
to a particular process in a given order can be determined 
by computing vacuum-polarization insertions in the diagrams of
that order.
The essence of the BLM procedure is that all vacuum-polarization 
effects from the QCD $\beta$ function are resummed into the
running coupling constant.

Let us just mention at this point that in addition to the BLM procedure,
two more renormalization scale-setting procedures in PQCD have been
proposed in the literature:
the principle of fastest apparent convergence (FAC) \cite{FAC},
and the principle of minimal sensitivity (PMS) \cite{PMS}.
The application of those three quite distinct methods can give 
strikingly different results in practical calculations
\cite{KramL91}.

As for the factorization scale $\nIR$,
it basically determines how much of the collinear term
given in \req{eq:fIR}, is absorbed into the distribution amplitude.
A natural choice for this scale would be
\Beq
     \nIR=Q^2 \, ,
\label{eq:nFQ}
\Eeq
which eliminates the logarithms of $Q^2/\nIR$.
More preferable to \req{eq:nFQ} is the choice
\Beq
   \nIR = \sqrt{\x \y} Q^2 \, ,
\label{eq:nFx}
\Eeq
which makes the function $f_{IR}$, given  by \req{eq:fIR},
vanish.

A glance at Eqs. (\ref{eq:Fpi}--\ref{eq:Fpi2b}), where 
the coupling constants
$\alpha_S(\nUV)$ and $\alpha_S(\nIR)$ appear under the
integral sign, reveals that any of the choices of $\mu_{R}$
given by (\ref{eq:nRg}--\ref{eq:nRge}),
and the choice of $\mu_{F}$ given by
\req{eq:nFx}, leads immediately to the problem
if the usual one-loop formula \req{eq:alphas}
for the effective QCD running coupling constant is employed.
Namely, then, regardless of how large $Q^2$ is,
the integration of Eqs. (\ref{eq:Fpi}--\ref{eq:Fpi2b})  
allows $\alpha_S(\nUV)$ to be evaluated near zero momentum transfer.
Two approaches are possible to circumvent this problem.
First, one can choose $\nUV$ and $\nIR$ to be effective constants
by taking $\nUV=\left< \nUV \right>$ and $\nIR=\left< \nIR \right>$,
respectively.
Second, one can introduce a cutoff in formula \req{eq:alphas}
with the aim of preventing the effective coupling from becoming infinite
for vanishing gluon momenta.

If the first approach is taken, 
Eqs. (\ref{eq:nRg}--\ref{eq:nRge}) and \req{eq:nFx} get replaced
by the averages
\Bml
\label{eq:averages}
\Beqa
  \nUV&=&\left< \x \y Q^2 \right> \, , 
\label{eq:anRg} \\
   \nUV&=&\sqrt{\left< \x \y Q^2 \right> \left< \y Q^2 \right>} \, ,
\label{eq:anRgq} \\
  \nUV&=&\left< e^{-5/3} \x \y Q^2 \right> \, , 
\label{eq:anRge}
\Eeqa
\Eml
and
\Beq
    \nIR = \sqrt{\left< \x  \y \right>} Q^2 
         \, ,
\label{eq:snF}
\Eeq
respectively.
Taking into account the fact that
$\left< \x \y \right>=\left< \x \right> \left< \y \right>$
and $\left< \x \right>=\left< \y \right>$,
it is possible to write Eqs. \req{eq:averages} and \req{eq:snF} 
in the respective forms:
\Bml
\label{eq:nR}
\Beqa
    \nUV&=&\left< \x \right>^{2} Q^2 \, , 
\label{eq:nRx2} \\
    \nUV&=&\left< \x \right>^{3/2} Q^2 \, ,
\label{eq:nRx3} \\
    \nUV&=&e^{-5/3} \left< \x \right>^{2} Q^2 \, , 
\label{eq:nRx2e}
\Eeqa
\Eml
and
\Beq
    \nIR = \left< \x \right> Q^2 \, .
\label{eq:snIR}
\Eeq

\widetext

The key quantity in the above considerations
is $\left< \x \right>$, the average value of the momentum fraction.
It depends on the form of the distribution amplitude,
and there is no unique way of defining it.
A possible definition is
\Beq
    \left< \x \right>(\nIR) = 
               \frac{\displaystyle \int_0^1 dx \, \phi(x,\nIR)\, \x }{
	             \displaystyle \int_0^1 dx \, \phi(x,\nIR) } 
                     \, .
\label{eq:sx1}
\Eeq
Owing to the fact that all distribution amplitudes under consideration
are centered around the value $x=0.5$, it follows trivially
from \req{eq:sx1} that for all of them
\Beq
  \left< \x \right>(\nIR)=
  0.5 \, .
\label{eq:vsx1}
\Eeq
An alternative way of defining $\left< \x \right>$,
motivated by the form of the LO expression for the pion
form factor \req{eq:Q2Fpi1old}, is
\Beq
   \left< \x \right>(\nIR)=
         \frac{\displaystyle \int_0^1 dx \, 
	        \displaystyle \frac{\phi^{LO}(x,\nIR)}{\x} \, \x}%
              {\displaystyle \int_0^1 dx \, 
	        \displaystyle \frac{\phi^{LO}(x,\nIR)}{\x}} 
		=
	 \frac{1}{3[1+B_2^{LO}(\nIR)+B_4^{LO}(\nIR)+\ldots]} 
	         \, .
\label{eq:sx3}
\Eeq
It should be noted, however, that this formula 
can be straightforwardly applied only 
if $\nIR=Q^2$. 
On the other hand, if instead of $\nIR=Q^2$ one chooses the factorization
scale to be as given by
\req{eq:snIR}, then Eqs. \req{eq:snIR} and
\req{eq:sx3} form a nontrivial system of simultaneous equations.
According to \req{eq:sx3}, one obtains
$\left< \x \right>_{as}(Q^2)=1/3$ for any $Q^2$, while 
$0.242 \leq \left< \x \right>_{CZ}(Q^2) \leq 0.262$ for
$4 \mbox{GeV}^2 \leq Q^2 \leq 100 \mbox{GeV}^2$
(similar values are obtained for $\phi_{P2}$ and $\phi_{P3}$).

When using the $\phi_{as}(x,\nIR)$ distribution
it appears reasonable to take
$\left < \x \right >(\nIR)$= ${\left < \x \right >}_{as}$=1/2.
This can be justified on the grounds that 
this distribution is concentrated around
$x=0.5$,
and is characterized by a very weak evolution.
On the other hand, 
for the end-point concentrated distributions
$\phi_{CZ}(x,\nIR)$, $\phi_{P2}(x,\nIR)$ and
$\phi_{P3}(x,\nIR)$, which exibit sizable
evolutional effects, it is more appropriate to take 
${\left < \x  \right >}(\nIR)$ as given by Eq. \req{eq:sx3}.

\narrowtext

As stated above, the divergence of the effective QCD coupling
$\alpha_S(\nUV)$, as given by \req{eq:alphas}, 
is the reason that it is not possible
to use the choices of $\nUV$ given by (\ref{eq:nRg}--\ref{eq:nRge})
and $\nIR$ given by \req{eq:nFx}.
Equation \req{eq:alphas} does not represent the nonperturbative
behavior 
of $\alpha_S(\nUV)$ for small $\nUV$, and a number of proposals have
been suggested for the form of the coupling constant in
this regime \cite{PaP80etc}.
The most exploited parameterization of the effective QCD coupling constant
at low energies has the form
\Beq
    \alpha_{S}(\nUV)=
  \frac{4 \pi}{\beta_{0} \ln 
    \left( \displaystyle\frac{\nUV+C^2}{\Lambda^2} \right)}
             \, ,
\label{eq:alphaSmod}
\Eeq
where the constant $C$ encodes the nonperturbative dynamics and is 
usually interpreted as an ``effective dynamical gluon mass'' $m_g$. 
For $\nUV \gg  C^2$, the effective coupling in \req{eq:alphaSmod}
coincides with the one-loop formula \req{eq:alphas},
whereas at low momentum transfer this formula ``freezes'' to some
finite but not necessarily small value.

In view of the confinement phenomenon, the modification \req{eq:alphaSmod} 
is very natural: the lower bound on the particle momenta is set by the 
inverse of the confinement radius. 
This is equivalent to a strong suppression for the 
propagation of particles with small momenta. Thus, in a consistent 
calculation in which \req{eq:alphaSmod} is assumed a modified gluon 
propagator should be used: 
\Beq
   \frac{1}{k^2} \rightarrow \frac{1}{k^2 - m_g^2} \, .
\label{eq:modprop}
\Eeq
However, if one attempts to calculate the LO prediction for the pion form factor
making use of \req{eq:alphaSmod} and \req{eq:modprop}, and taking 
$m_g=300 - 500$ MeV, the result obtained is by a 
factor of $10$ lower than the experimental value, questioning the 
applicability of such an approach \cite{Rad91}.

It has recently been shown that spontaneous chiral symmetry breaking
imposes rather a severe restriction on the idea of freezing \cite{PeR97}. 
The authors of \cite{PeR97} argue that before any argument based on 
a particular 
form of the freezing coupling constant is put forward, one should check 
that the dynamical origin (mechanism) of the freezing is such that 
enough chiral symmetry breaking can be produced. 

Considering the discussion above, calculations with the 
frozen coupling constant seem to need a more refined treatment and will 
not be considered in this paper.

\section{Complete NLO numerical predictions for the pion form factor}

Having obtained all the necessary
ingredients in the preceding sections, now we put them together and 
obtain complete
leading-twist NLO QCD numerical predictions for the pion
form factor.
For a fixed distribution amplitude, 
we analyze the dependence of our results on the choice
of the scales $\mu_R$ and $\mu_F$.

\widetext

By inserting 
\req{eq:phiLOexp} and \req{eq:phiNLOexp} into
\req{eq:Fpi1} and \req{eq:Fpi2b}, 
taking into account \req{eq:TH0}, 
taking the scales
$\nUV$ and $\nIR$ to be effective constants
(as explained in Sec. IV),
and performing the $x$ and $y$ integration,
we find that Eqs.
\req{eq:Fpi1} and \req{eq:Fpi2b} 
take the form
\Beq
  Q^2 F_{\pi}^{(0)}(Q^2,\nUV,\nIR)  =  8 \pi f_{\pi}^2 
                \alpha_S(\nUV)
                \left( \sum_{n=0}^{\infty}{}' B_n^{LO}(\nIR) \right)^2
		 \, ,
\label{eq:Q2Fpi0gen}
\Eeq
and
\Beq
  Q^2 F_{\pi}^{(1b)}(Q^2,\nUV,\nIR)  =  16 f_{\pi}^2 
                \alpha_S(\nUV) \alpha_S(\nIR)
                \left( \sum_{n=0}^{\infty}{}' B_n^{LO}(\nIR) \right)
                \left( \sum_{k=2}^{\infty}{}' B_k^{NLO}(\nIR) \right)
		\, .
\label{eq:Q2Fpi1bgen}
\Eeq
For a distribution amplitude of the form given in
(\ref{eq:phi4exp} - \ref{eq:BnNLO4}),
the above expression reduces to
\Beqa
  Q^2 F_{\pi}^{(0)}(Q^2,\nUV,\nIR) & = & 8 \, \pi \, f_{\pi}^2 \,  
                \alpha_S(\nUV) \,
                \left[ 1 + B_2^{LO}(\nIR) + B_4^{LO}(\nIR) \right] ^2
		   \, ,
\label{eq:Q2Fpi0}
		\\
  Q^2 F_{\pi}^{(1b)}(Q^2,\nUV,\nIR) & = & 16 \, f_{\pi}^2  \,
                \alpha_S(\nUV)  \, \alpha_S(\nIR) \,
                \left[ 1 + B_2^{LO}(\nIR) + B_4^{LO}(\nIR) \right] \, 
		 \nonumber \\ & &  \times
                \left[ B_2^{NLO}(\nIR) + B_4^{NLO}(\nIR) 
		         + \sum_{k=6}^{\infty} {}' B_k^{NLO}(\nIR) \right] 
		\, .
\label{eq:Q2Fpi1b}
\Eeqa
As for the part of the NLO contribution arising from the NLO correction
to the hard-scattering amplitude, by
inserting \req{eq:TH1} and \req{eq:phiLO4exp} into \req{eq:Fpi2a}
and performing the $x$ and $y$ integration one obtains
\Beqa
  Q^2 F_{\pi}^{(1a)}(Q^2,\nUV,\nIR) & = & 8 \,  f_{\pi}^2  \,
                \alpha_S^2(\nUV) \,
                \left\{
		   \frac{9}{4} 
                \left[ 1 + B_2^{LO}(\nIR) + B_4^{LO}(\nIR) \right] ^2
                   \ln \frac{\nUV}{Q^2} 
		\right. \nonumber \\ & & \left.
		  + \frac{2}{3} 
                \left[ \frac{25}{6} B_2^{LO}(\nIR) + 
		       \frac{91}{15} B_4^{LO}(\nIR) \right]
                \left[ 1 + B_2^{LO}(\nIR) + B_4^{LO}(\nIR) \right]
                   \ln \frac{\nIR}{Q^2}
		\right. \nonumber \\ & & \left.
		  + 6.58 
		   + 24.99 \, B_2^{LO}(\nIR)+ 21.43 \, (B_2^{LO}(\nIR))^2 
		\right. \nonumber \\ & & \left.
		   + 32.81 \, B_4^{LO}(\nIR)+ 32.55 \, (B_4^{LO}(\nIR))^2
		   + 53.37 \, B_2^{LO}(\nIR) \, B_4^{LO}(\nIR) \right\} 
		    \, .
\label{eq:Q2Fpi1a}
\Eeqa
Then the NLO contribution to the ``almost scaling'' combination
$ Q^2 F_{\pi}(Q^2)$ is given by
\Beq
   Q^2 F_{\pi}^{(1)}(Q^2,\nUV,\nIR) = Q^2 F_{\pi}^{(1a)}(Q^2,\nUV,\nIR) +
                                Q^2 F_{\pi}^{(1b)}(Q^2,\nUV,\nIR)
				\, ,
\label{eq:Q2Fpi1}
\Eeq
while the total NLO prediction reads
\Beq
   Q^2 F_{\pi}(Q^2,\nUV,\nIR) = Q^2 F_{\pi}^{(0)}(Q^2,\nUV,\nIR) +
                                Q^2 F_{\pi}^{(1)}(Q^2,\nUV,\nIR)
				\, .
\label{eq:Q2Fpi}
\Eeq

\narrowtext

For the purpose of this calculation we adopt the criteria
according to which
a perturbative prediction for $F_{\pi}(Q^2)$ 
is considered reliable provided the following two requirements
are met:
first, corrections to the LO prediction are reasonably small
($< 30\%$);
second, the expansion parameter (effective QCD coupling constant) is
acceptably small ($\alpha_S(\mu_R^2)< 0.3$ or $0.5$).
Of course, one more requirement should be added 
to the above ones: consistency with experimental data.
This requirement, however, is not of much use here since reliable
experimental data
for the pion form factor exist for $Q^2 \leq 4$ GeV$^2$, i.e., well
outside the region in which the perturbative treatment based on
Eq. \req{eq:piffcf} is justified.

\begin{figure}
\centerline{\epsfig{file=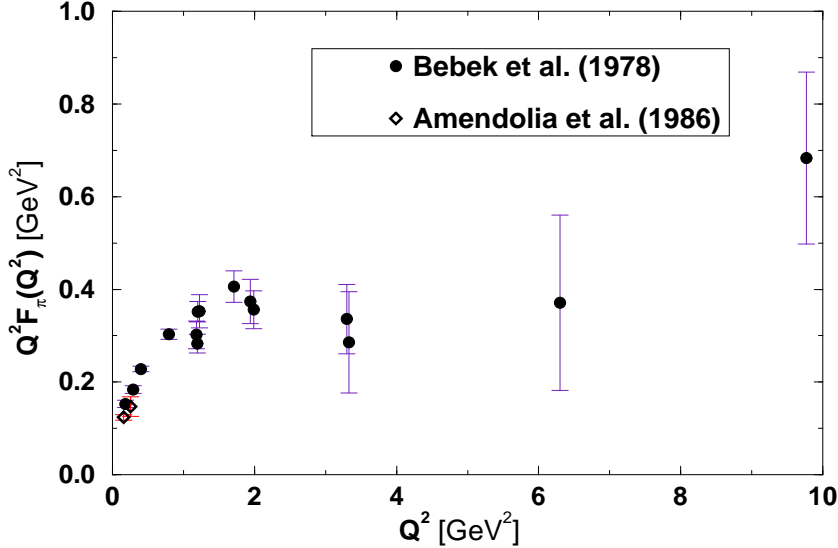,height=8cm,width=11cm,silent=}}
\caption{Presently available experimental data for the
         spacelike pion electromagnetic form factor.}
\label{f:exp}
\end{figure}

Currently available experimental data for the spacelike
pion electromagnetic form factor $F_{\pi}(Q^2)$ are
shown in Fig. \ref{f:exp}.
The data are taken from Bebek et al. \cite{Be78} and
Amendolia et al. \cite{Am86}.
As stated in \cite{Be78}, the measurements corresponding to
$Q^2=6.30$GeV$^2$ and $Q^2=9.77$GeV$^2$ are somewhat
questionable. 
Thus, effectively, the data for
$F_{\pi}(Q^2)$ exist only for $Q^2$ in the range 
$Q^2 \leq 4$GeV$^2$.
The results of Ref. \cite{Be78} were obtained from
the extrapolation of the 
$\gamma^* p \rightarrow \pi^+ n$ electroproduction data
to the pion pole.
It should also be mentioned that the analysis of Ref. \cite{Be78}
is subject to criticism which questions whether $F_{\pi}(Q^2)$
was truly determined for $1 \leq Q^2 \leq 4$GeV$^2$ \cite{CaM90}
(but see also Ref. \cite{DubM89etc}).
The new data in this energy region are expected from the CEBAF experiment
E-93-021.

\subsection{Predictions obtained with $\nUV=\nIR=Q^2$}
\label{ssec:nrQ}

\narrowtext

The first NLO prediction for the pion form factor was obtained 
in Ref. \cite{FiG81}.
Using the $\overline{MS}$ renormalization scheme and the choice 
$\nUV=\nIR=Q^2$ it was found that for the $\phi_{as}(x)$ distribution
(with the evolution of the distribution amplitude neglected),
the perturbative series took the form
\Beq
      Q^2 F_{\pi}(Q^2) = ( 0.43 \mbox{ GeV}^2) \, \alpha_S(Q^2)
            \, [ 1 + 2.10 \, \alpha_S(Q^2) + \cdots ] 
	    \, ,
\Eeq
which is in agreement with our result given by
\req{eq:Q2Fpi}, \req{eq:Q2Fpi0}, and \req{eq:Q2Fpi1a}.
The conclusion based on this result was that a reliable result
for $F_{\pi}(Q^2)$ was not obtained until
$\alpha_S(Q^2) \approx 0.1$, or with 
$\Lambda_{\overline{MS}}=0.5$ GeV, $Q^2=10000$ GeV$^2$.
This prediction has been widely cited in the literature and
initiated a lot of discussion regarding the applicability of
PQCD to the calculation of exclusive processes 
at large momentum transfer.
With the presently accepted value of 0.2 GeV for 
$\Lambda_ {\overline{MS}}$, we find that the criteria from Ref. \cite{FiG81}
are satisfied for $Q^2\sim 1600$ GeV$^2$,
 and that for $Q^2=700$ GeV$^2$, the corrections to the 
LO prediction are of order 30\%.
Thus, this result shows that for the choice of
the renormalization and factorization scales $\nUV=\nIR=Q^2$,
the region in which perturbative predictions can be considered
reliable is still well beyond the region in which experimental data exist.
The inclusion of the distribution amplitude evolution effects,
although extremely important for the end-point concentrated amplitudes,
does not change this conclusion.

\begin{figure}
\centerline{\epsfig{file=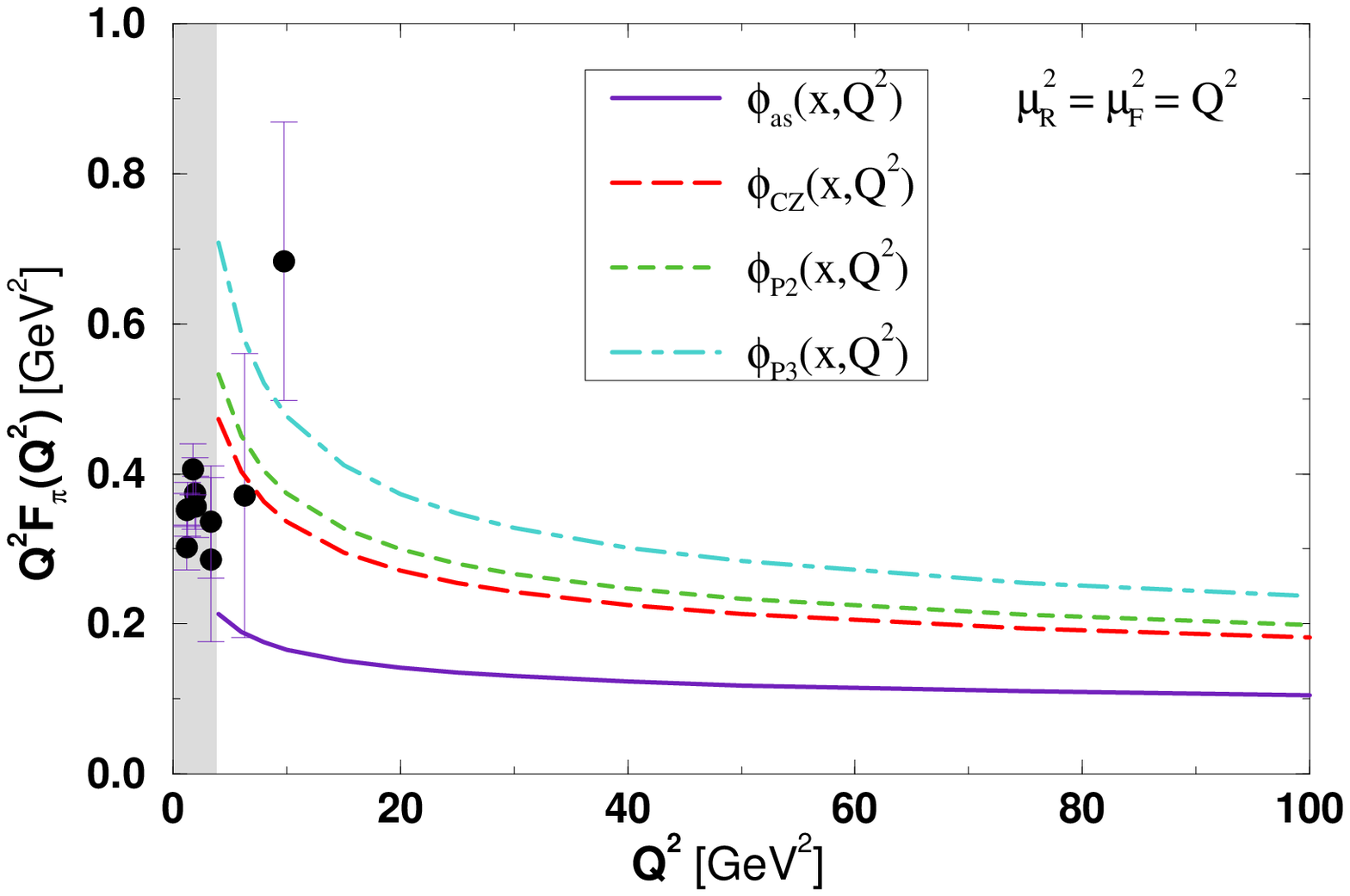,height=8cm,width=11cm,silent=}}
\caption{Comparison of the complete leading-twist NLO QCD predictions for the 
    pion form factor, $Q^2 F_{\pi}(Q^2)$, obtained using the four
    candidate distribution amplitudes, 
    with the presently available experimental
    data.
    The shaded area denotes the region of $Q^2$ in which 
    $\alpha_S(Q^2)\!>\!0.3$.}
\label{f:Aa}
%
\centerline{\epsfig{file=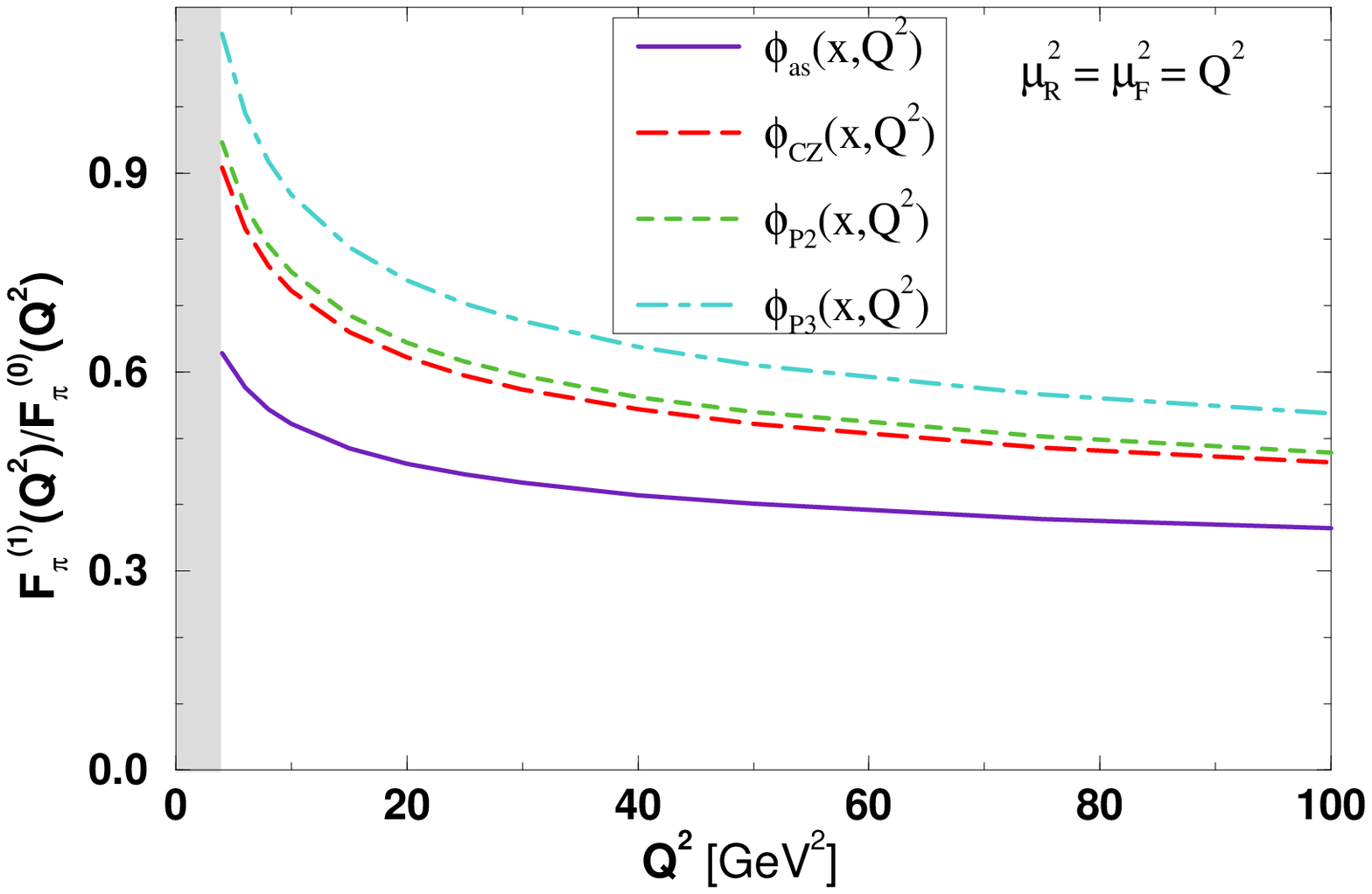,height=8cm,width=11cm,silent=}}
\caption{The ratio of the NLO to the
    LO contributions to the pion form factor,
    $F_{\pi}^{(1)}(Q^2)/F_{\pi}^{(0)}(Q^2)$,
    for the four candidate distribution amplitudes.
    The shaded area denotes the region of $Q^2$ in which 
    $\alpha_S(Q^2)\!>\!0.3$.}
\label{f:ratioAa}
\end{figure}

Numerical results of our complete NLO QCD calculation,
obtained using the four candidate distribution 
amplitudes, with $\nUV=\nIR=Q^2$, and for $Q^2 \geq 4$ GeV$^2$, 
are displayed in Tables \ref{t:rASaA}, \ref{t:rCZaA},
\ref{t:rP2aA}, and \ref{t:rP3aA}.
The entries in these tables include various contributions given by
Eqs. (\ref{eq:Q2Fpi0}--\ref{eq:Q2Fpi}), 
comprising the full NLO result.
A comparison of our results with presently available experimental data 
is shown in Fig. \ref{f:Aa}.
The ratio of the NLO to the LO contribution to 
$F_{\pi}(Q^2)$, i.e., $F_{\pi}^{(1)}(Q^2)/F_{\pi}^{(0)}(Q^2)$, 
as a useful measure of the importance of the NLO corrections,
 is plotted as a function of $Q^2$ in Fig. \ref{f:ratioAa}.
The shaded areas appearing in Figs. \ref{f:Aa} and \ref{f:ratioAa}
denote the region of $Q^2$ where $\alpha_S(Q^2) > 0.3$.
We take that outside of this region the effective coupling 
is acceptably small.

\mediumtext

\begin{table}
\caption{Complete leading-twist NLO QCD results for the pion form factor,
         $Q^2 F_{\pi}(Q^2)$, obtained using the
         $\phi_{as}(x,\nIR)$ distribution amplitude and
         assuming $\nUV=\nIR=Q^2$.}

\begin{tabular}{cccccccc} 
$Q^2$ & $\alpha_S(\nUV)$ & $Q^2 F_{\pi}^{(0)}(Q^2)$ &
$Q^2 F_{\pi}^{(1a)}(Q^2)$ & $Q^2 F_{\pi}^{(1b)}(Q^2)$ &
$Q^2 F_{\pi}^{(1)}(Q^2)$ & 
$F_{\pi}^{(1)}(Q^2)/F_{\pi}^{(0)}(Q^2)$ & 
$Q^2 F_{\pi}(Q^2)$ \\ 
$[\mbox{GeV}^{2}]$ & & $[\mbox{GeV}^{2}]$ & 
$[\mbox{GeV}^{2}]$ & $[\mbox{GeV}^{2}]$ & $[\mbox{GeV}^{2}]$ &
 \% & $[\mbox{GeV}^{2}]$ 
 \\[.05cm] \hline  
  4 & 0.303 & 0.131 & 0.083 & -0.001 & 0.082 & 62.8 & 0.213\\ 
  6 & 0.279 & 0.120 & 0.070 & -0.001 & 0.069 & 57.6 & 0.189\\
  8 & 0.264 & 0.114 & 0.063 & -0.001 & 0.062 & 54.4 & 0.176\\
 10 & 0.253 & 0.109 & 0.058 & -0.001 & 0.057 & 52.2 & 0.166\\
 20 & 0.225 & 0.097 & 0.046 & -0.001 & 0.045 & 46.2 & 0.142\\
 30 & 0.211 & 0.091 & 0.040 & -0.001 & 0.039 & 43.3 & 0.130\\
 40 & 0.202 & 0.087 & 0.037 & -0.001 & 0.036 & 41.4 & 0.123\\
 50 & 0.196 & 0.085 & 0.035 & -0.001 & 0.034 & 40.1 & 0.118\\
 75 & 0.185 & 0.080 & 0.031 & -0.001 & 0.030 & 37.9 & 0.110\\
100 & 0.178 & 0.077 & 0.029 & -0.001 & 0.028 & 36.4 & 0.105\\
\end{tabular}

\label{t:rASaA}
\end{table}
\begin{table}

\caption{Same as Table \protect\ref{t:rASaA} but for
         the $\phi_{CZ}(x,\nIR)$ distribution amplitude.}

\begin{tabular}{cccccccc} 
$Q^2$ & $\alpha_S(\nUV)$ & $Q^2 F_{\pi}^{(0)}(Q^2)$ &
$Q^2 F_{\pi}^{(1a)}(Q^2)$ & $Q^2 F_{\pi}^{(1b)}(Q^2)$ &
$Q^2 F_{\pi}^{(1)}(Q^2)$ & 
$F_{\pi}^{(1)}(Q^2)/F_{\pi}^{(0)}(Q^2)$ & 
$Q^2 F_{\pi}(Q^2)$ \\ 
$[\mbox{GeV}^{2}]$ & & $[\mbox{GeV}^{2}]$ & 
$[\mbox{GeV}^{2}]$ & $[\mbox{GeV}^{2}]$ & $[\mbox{GeV}^{2}]$ &
 \% & $[\mbox{GeV}^{2}]$ 
 \\[.05cm] \hline  
  4 & 0.303 & 0.248 & 0.241 & -0.015 & 0.225 & 90.8 & 0.474 \\ 
  6 & 0.279 & 0.222 & 0.195 & -0.014 & 0.181 & 81.6 & 0.403 \\
  8 & 0.264 & 0.206 & 0.170 & -0.013 & 0.157 & 76.1 & 0.363 \\
 10 & 0.253 & 0.195 & 0.153 & -0.012 & 0.141 & 72.2 & 0.336 \\
 20 & 0.225 & 0.167 & 0.115 & -0.011 & 0.104 & 62.2 & 0.271 \\
 30 & 0.211 & 0.154 & 0.098 & -0.010 & 0.088 & 57.4 & 0.243 \\
 40 & 0.202 & 0.146 & 0.089 & -0.009 & 0.079 & 54.3 & 0.225 \\
 50 & 0.196 & 0.140 & 0.082 & -0.009 & 0.073 & 52.2 & 0.213 \\
 75 & 0.185 & 0.131 & 0.072 & -0.008 & 0.064 & 48.7 & 0.194 \\
100 & 0.178 & 0.125 & 0.065 & -0.008 & 0.058 & 46.4 & 0.182 \\
\end{tabular}

\label{t:rCZaA}
\end{table}
\begin{table}

\caption{Same as Table \protect\ref{t:rASaA} but for
         the $\phi_{P2}(x,\nIR)$ distribution amplitude.}

\begin{tabular}{cccccccc} 
$Q^2$ & $\alpha_S(\nUV)$ & $Q^2 F_{\pi}^{(0)}(Q^2)$ &
$Q^2 F_{\pi}^{(1a)}(Q^2)$ & $Q^2 F_{\pi}^{(1b)}(Q^2)$ &
$Q^2 F_{\pi}^{(1)}(Q^2)$ & 
$F_{\pi}^{(1)}(Q^2)/F_{\pi}^{(0)}(Q^2)$ & 
$Q^2 F_{\pi}(Q^2)$ \\ 
$[\mbox{GeV}^{2}]$ & & $[\mbox{GeV}^{2}]$ & 
$[\mbox{GeV}^{2}]$ & $[\mbox{GeV}^{2}]$ & $[\mbox{GeV}^{2}]$ &
 \%  & $[\mbox{GeV}^{2}]$ 
 \\[.05cm] \hline  
  4 & 0.303 & 0.274 & 0.278 & -0.019 & 0.259 & 94.6 & 0.532 \\ 
  6 & 0.279 & 0.244 & 0.224 & -0.017 & 0.207 & 85.0 & 0.451 \\
  8 & 0.264 & 0.226 & 0.195 & -0.016 & 0.179 & 79.1 & 0.404 \\
 10 & 0.253 & 0.213 & 0.175 & -0.015 & 0.160 & 75.0 & 0.374 \\
 20 & 0.225 & 0.182 & 0.130 & -0.013 & 0.117 & 64.5 & 0.300 \\
 30 & 0.211 & 0.167 & 0.111 & -0.012 & 0.100 & 59.4 & 0.267 \\
 40 & 0.202 & 0.158 & 0.100 & -0.011 & 0.089 & 56.2 & 0.247 \\
 50 & 0.196 & 0.152 & 0.093 & -0.011 & 0.082 & 54.0 & 0.234 \\
 75 & 0.185 & 0.141 & 0.081 & -0.010 & 0.071 & 50.2 & 0.212 \\
100 & 0.178 & 0.134 & 0.074 & -0.009 & 0.064 & 47.9 & 0.199 \\
\end{tabular}

\label{t:rP2aA}
\end{table}
\begin{table}

\caption{Same as Table \protect\ref{t:rASaA} but for
         the $\phi_{P3}(x,\nIR)$ distribution amplitude.}

\begin{tabular}{cccccccc} 
$Q^2$ & $\alpha_S(\nUV)$ & $Q^2 F_{\pi}^{(0)}(Q^2)$ &
$Q^2 F_{\pi}^{(1a)}(Q^2)$ & $Q^2 F_{\pi}^{(1b)}(Q^2)$ &
$Q^2 F_{\pi}^{(1)}(Q^2)$ & 
$F_{\pi}^{(1)}(Q^2)/F_{\pi}^{(0)}(Q^2)$ & 
$Q^2 F_{\pi}(Q^2)$ \\ 
$[\mbox{GeV}^{2}]$ & & $[\mbox{GeV}^{2}]$ & 
$[\mbox{GeV}^{2}]$ & $[\mbox{GeV}^{2}]$ & $[\mbox{GeV}^{2}]$ &
\%  & $[\mbox{GeV}^{2}]$ 
 \\[.05cm] \hline  
  4 & 0.303 & 0.335 & 0.402 & -0.029 & 0.372 & 111.0 & 0.708 \\ 
  6 & 0.279 & 0.295 & 0.318 & -0.026 & 0.292 & 100.0 & 0.587 \\
  8 & 0.264 & 0.272 & 0.273 & -0.024 & 0.249 &  91.8 & 0.521 \\
 10 & 0.253 & 0.255 & 0.244 & -0.023 & 0.222 &  86.7 & 0.477 \\
 20 & 0.225 & 0.215 & 0.178 & -0.019 & 0.158 &  73.8 & 0.373 \\
 30 & 0.211 & 0.196 & 0.150 & -0.017 & 0.133 &  67.7 & 0.328 \\
 40 & 0.202 & 0.184 & 0.134 & -0.016 & 0.118 &  63.8 & 0.302 \\
 50 & 0.196 & 0.176 & 0.123 & -0.015 & 0.108 &  61.1 & 0.284 \\
 75 & 0.185 & 0.163 & 0.106 & -0.014 & 0.092 &  56.6 & 0.255 \\
100 & 0.178 & 0.154 & 0.096 & -0.013 & 0.083 &  53.7 & 0.237 \\
\end{tabular}

\label{t:rP3aA}
\end{table}

\narrowtext

It is evident from Figs. \ref{f:Aa} and \ref{f:ratioAa}
and Tables \ref{t:rASaA}--\ref{t:rP3aA}
that the leading-twist NLO results for the pion form factor
obtained with $\nUV=\nIR=Q^2$ display the following general features.
First, the results are quite sensitive
to the assumed form of the pion distribution amplitude.
Thus, the more end-point concentrated distribution amplitude is,
the larger result for the pion form factor is obtained,
and also the NLO corrections are larger
(which has already been obvious looking at  Eqs. %
(\ref{eq:Q2Fpi0}-\ref{eq:Q2Fpi})).
Second, whereas the NLO correction arising from the corrections 
to the hard-scattering amplitude are positive, 
the corrections due to the inclusion of the evolutional
corrections to the distribution amplitude are negative, with the former
being generally an order of magnitude larger than the latter. 
Thus, in all the cases considered, the full NLO correction to the
pion form factor is positive, i.e., its inclusion increases the LO
prediction.

We now briefly comment on the results obtained with each of the
four distribution amplitudes.

Table \ref{t:rASaA}, which corresponds to the
$\phi_{as}(x,Q^2)$ distribution amplitude, 
shows that the NLO correction
$Q^2 F_{\pi}^{(1)}(Q^2)$ is rather large (36\% at $Q^2=100$ GeV$^2$).
Most of the contribution to $Q^2 F_{\pi}^{(1)}(Q^2)$ is due to the
NLO correction to the hard-scattering amplitude
$Q^2 F_{\pi}^{(1a)}(Q^2)$, while the contribution
$Q^2 F_{\pi}^{(1b)}(Q^2)$ arising from the NLO evolutional correction
of the distribution amplitude is rather small,
being of order 1\%.
The ratio $F_{\pi}^{(1)}(Q^2)/F_{\pi}^{(0)}(Q^2) \approx 30\%$
is reached at $Q^2 \approx 500$ GeV$^2$.

\begin{figure}
\centerline{\epsfig{file=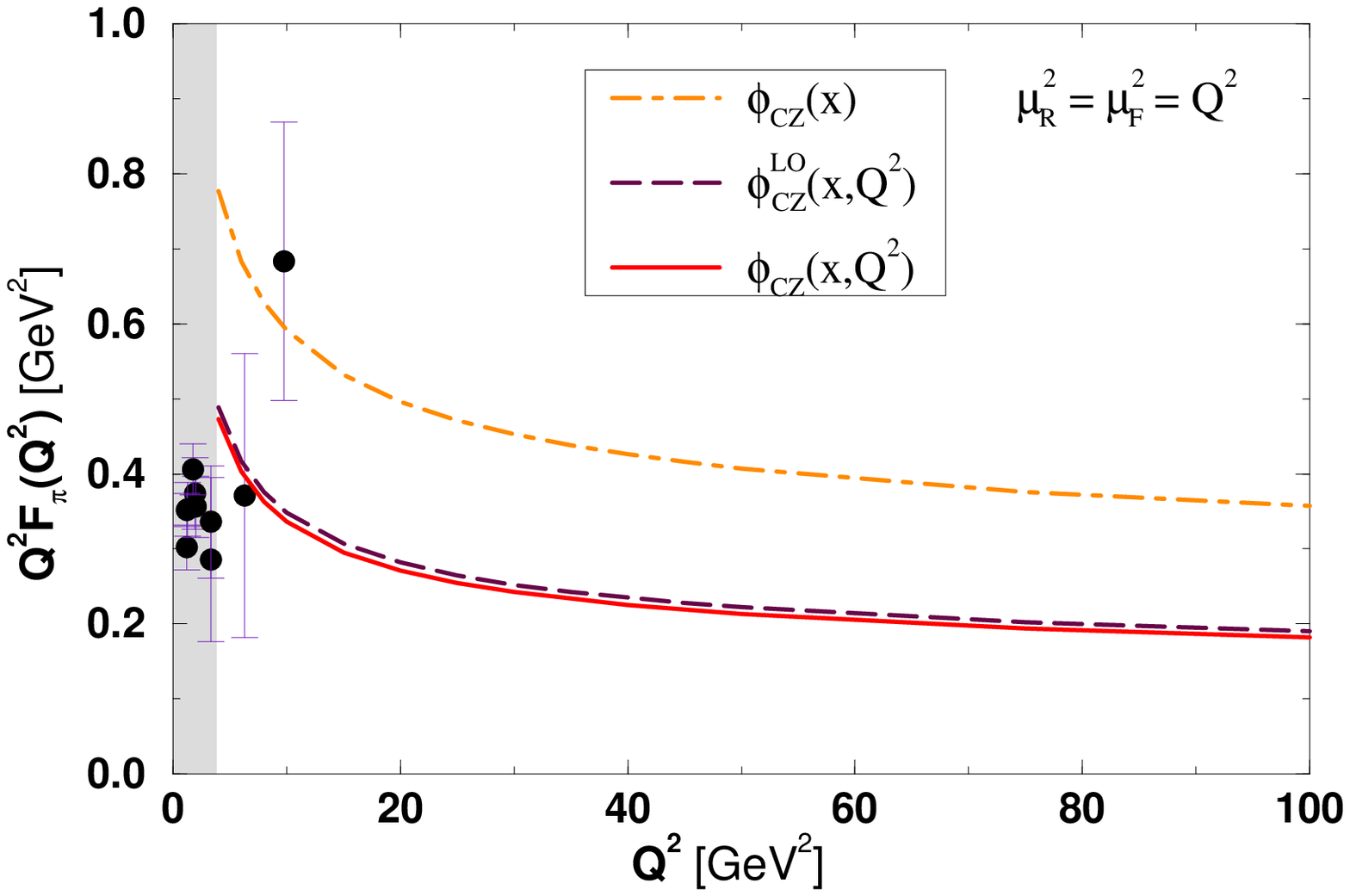,height=8cm,width=11cm,silent=}}
\caption{Leading-twist NLO QCD predictions for the pion form factor,
    $Q^2 F_{\pi}(Q^2)$, obtained with the
    $\phi_{CZ}(x)$, $\phi_{CZ}^{LO}(x,Q^2)$,
     and $\phi_{CZ}(x,Q^2)$ distributions. 
    The shaded area denotes the region of $Q^2$ in which 
    $\alpha_S(Q^2)\!>\!0.3$.}
\label{f:CZaA}
%
\centerline{\epsfig{file=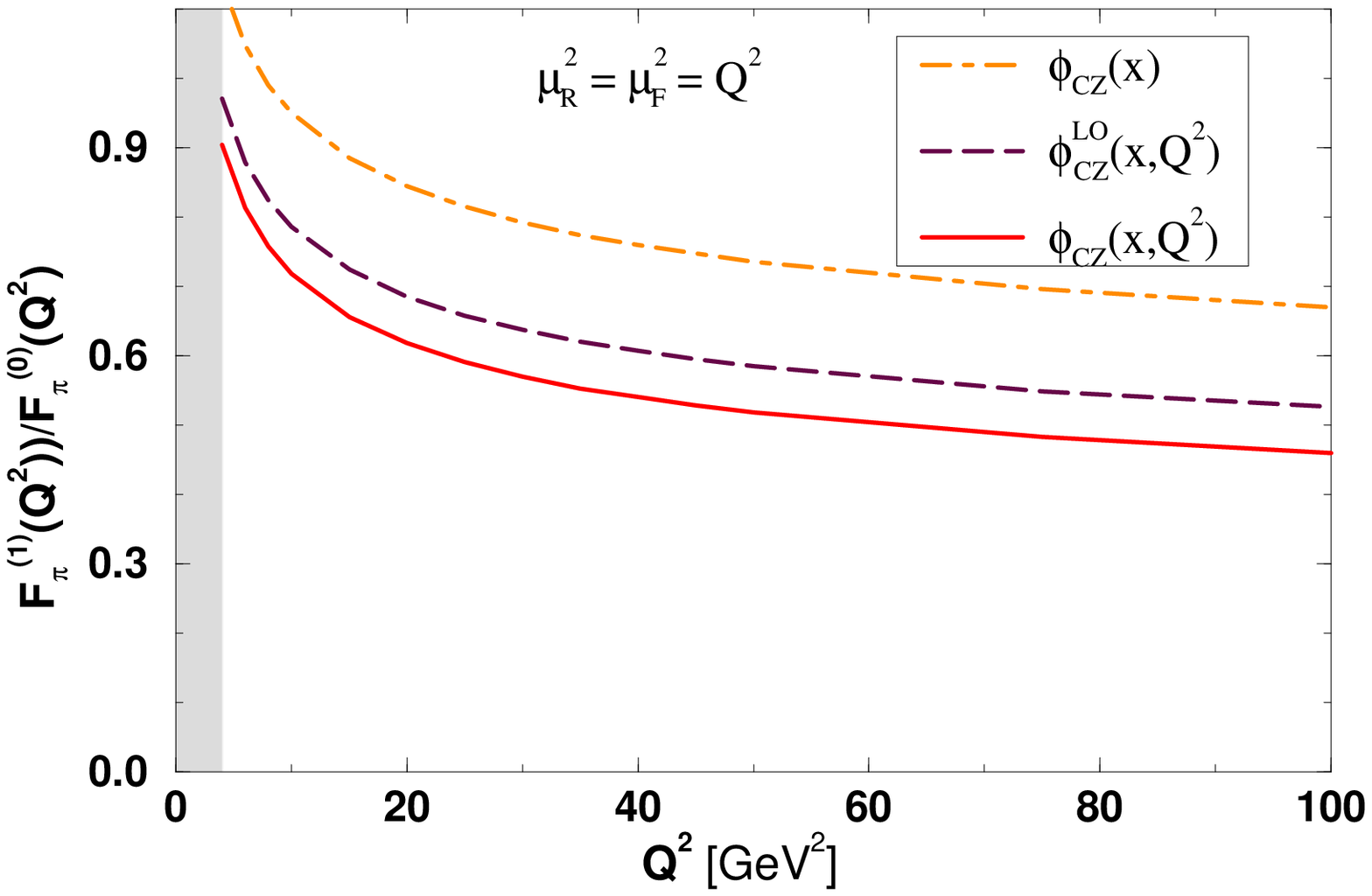,height=8cm,width=11cm,silent=}}
\caption{The ratio of the NLO to the
     LO contributions to the pion form factor,
    $F_{\pi}^{(1)}(Q^2)/F_{\pi}^{(0)}(Q^2)$,
    obtained for the $\phi_{CZ}(x)$, $\phi_{CZ}^{LO}(x,Q^2)$,
     and $\phi_{CZ}(x,Q^2)$ distributions.
    The shaded area denotes the region of $Q^2$ in which 
    $\alpha_S(Q^2)\!>\!0.3$.}
\label{f:ratioCZaA}
\end{figure}

The results derived from the $\phi_{CZ}(x,Q^2)$ distribution
are presented in Table \ref{t:rCZaA}.
The full NLO correction $Q^2 F_{\pi}^{(1)}(Q^2)$
is somewhat larger than for $\phi_{as}(x,Q^2)$ 
(and at $Q^2=100$ GeV$^2$ it amounts to 46\%).
The ratio $F_{\pi}^{(1)}(Q^2)/F_{\pi}^{(0)}(Q^2)$ is greater than
$30\%$
until $Q^2 \approx 2400$ GeV$^2$.
It is important to observe that the evolutional corrections,
especially the LO ones,
are rather significant in this case.
Also, as it is seen from Table \ref{t:rCZaA}, the NLO evolutional correction
$Q^2 F_{\pi}^{(1b)}(Q^2)$ is of order $\approx 6 \%$,
i.e., nonnegligible.
To show the importance of the correction arising from the inclusion
of the distribution amplitude evolution, the results
for $Q^2 F_{\pi}(Q^2)$
and the ratio $F_{\pi}^{(1)}(Q^2)/F_{\pi}^{(0)}(Q^2)$
obtained using the $\phi_{CZ}(x)$, $\phi_{CZ}^{LO}(x,Q^2)$, and 
$\phi_{CZ}(x,Q^2)$ distributions, are exhibited
in Figs. \ref{f:CZaA} and \ref{f:ratioCZaA}, respectively.

The results based on the $\phi_{P2}(x,Q^2)$  and
$\phi_{P3}(x,\nIR)$ distributions
are listed in Table \ref{t:rP2aA} and \ref{t:rP3aA}, respectively.
As it can be easily seen by looking at Figs.
\ref{f:Aa} and \ref{f:ratioAa} and by comparing the corresponding
entries in Tables \ref{t:rCZaA} and \ref{t:rP2aA},
the results obtained with the $\phi_{P2}(x,Q^2)$ and
$\phi_{CZ}(x,Q^2)$ distributions are practically the same qualitatively, 
while differ quantitatively by a few percent.
From Table \ref{t:rP3aA} and Figs. \ref{f:Aa} and \ref{f:ratioAa},
one can see that
the behavior of the results 
obtained with the $\phi_{P3}(x,\nIR)$ distribution  is  
qualitatively similar to the behavior of the
results obtained with other two end-point concentrated distributions.
For this reason, we leave the $\phi_{P2}(x,Q^2)$ and
$\phi_{P3}(x,\nIR)$  distributions
out of our further consideration. 

%

In view of what has been said above, we may conclude the
following. 
If the pion is modeled by the
$\phi_{as}(x,\nIR)$, $\phi_{CZ}(x,\nIR)$,
$\phi_{P2}(x,\nIR)$, or $\phi_{P3}(x,\nIR)$
distribution amplitude, and if the renormalization and factorization
scale are chosen to be $\nUV=\nIR=Q^2$,
one finds that the NLO corrections to the lowest-order prediction
for the pion form factor are large.
The NLO predictions obtained cannot be made reliable,
i.e., $F_{\pi}^{(1)}(Q^2)/F_{\pi}^{(0)}(Q^2)$ less than, say,
30 \%, until the momentum transfer $Q \gg 10$ GeV is reached. 
Based on these findings and considering
the region of $Q^2$ in which the data exist, 
it is clear that we are not in a 
position to rule out any of the four distributions considered.
One can only note that the predictions for 
$Q^2 F_{\pi}(Q^2,\nUV,\nIR)$ obtained with the
$\phi_{as}(x,\nIR)$ distribution are below the trend indicated by the
existing experimental data, while the
end-point concentrated distributions $\phi_{CZ}(x,\nIR)$,
$\phi_{P2}(x,\nIR)$, and $\phi_{P3}(x,\nIR)$ give higher predictions.
It is worth mentioning here that 
the theoretical predictions for the photon--to--pion
transition form factor $F_{\pi \gamma}(Q^2)$ 
are in very good agreement with the data,
assuming the pion distribution amplitude is close to
the asymptotic one, i.e., $\phi_{as}(x,\nIR)$ \cite{Rad95etc}.

\subsection{Predictions obtained using
            $\nUV=a \, Q^2$ and $\nIR=b \, Q^2$}

In this subsection we present a detailed analysis of the dependence of the
complete leading$-$twist NLO predictions for the pion form factor obtained
with the
${\phi}_{as}(x,\nIR)$ and ${\phi}_{CZ}(x,\nIR)$ distributions, 
on the renormalization and factorization scales, ${\mu}_R$ and ${\mu}_F$.
In the following we shall restrict our attention to two most exploited
pion distribution amplitudes
${\phi}_{as}(x,\nIR)$ and ${\phi}_{CZ}(x,\nIR)$.

In the preceding subsection, we have found that the NLO corrections
calculated using these two distributions are large, especially for the latter.
The reason for this lies in the fact that the renormalization scale choice
$\nUV=Q^2$ is not appropriate one. 
Namely, owing to the partitioning of the
overall momentum transfer $Q^2$ among the particles in the parton subprocess,
the essential virtualities of the particles are smaller than $Q^2$, so that
the physical renormalization scale, better suited for analyzing the process
under consideration, is inevitably lower than the external scale $Q^2$.

Compared with the scale $\mu_R$, the factorization scale
$\mu_F$ turns out to be of secondary importance.

A characteristic feature of the asymptotic distribution
${\phi}_{as}(x,\nIR)$ is that in the LO it shows no evolution.
A consequence of this, 
as it can be seen from Eqs. (\ref{eq:Q2Fpi0}-\ref{eq:Q2Fpi}), 
is that the NLO predictions for the pion form factor
based on this distribution are essentially independent of the
factorization scale $\mu_F$.
Namely, the only dependence on this scale is contained in the
term arising from the NLO evolutional effects, which 
are tiny, as we have seen in Sec. III.
On the other hand, the predictions calculated with the
${\phi}_{CZ}(x,\nIR)$
are ${\mu}_F$ dependent, but this dependence turns out to be very weak.
Thus, for a given value of
$\mu_R$, variation of the value of $\mu_F$ in the range
$Q^2/4 \leq \nIR \leq Q^2$ leads to practically the same results.
Therefore, 
when using the ${\phi}_{as}(x,\nIR)$ and ${\phi}_{CZ}(x,\nIR)$ distributions,  
one is allowed to set $\nIR=Q^2$, for all practical purposes.

\subsubsection{Examining the renormalization scale dependence of the
               NLO corrections}

\begin{figure}
\centerline{\epsfig{file=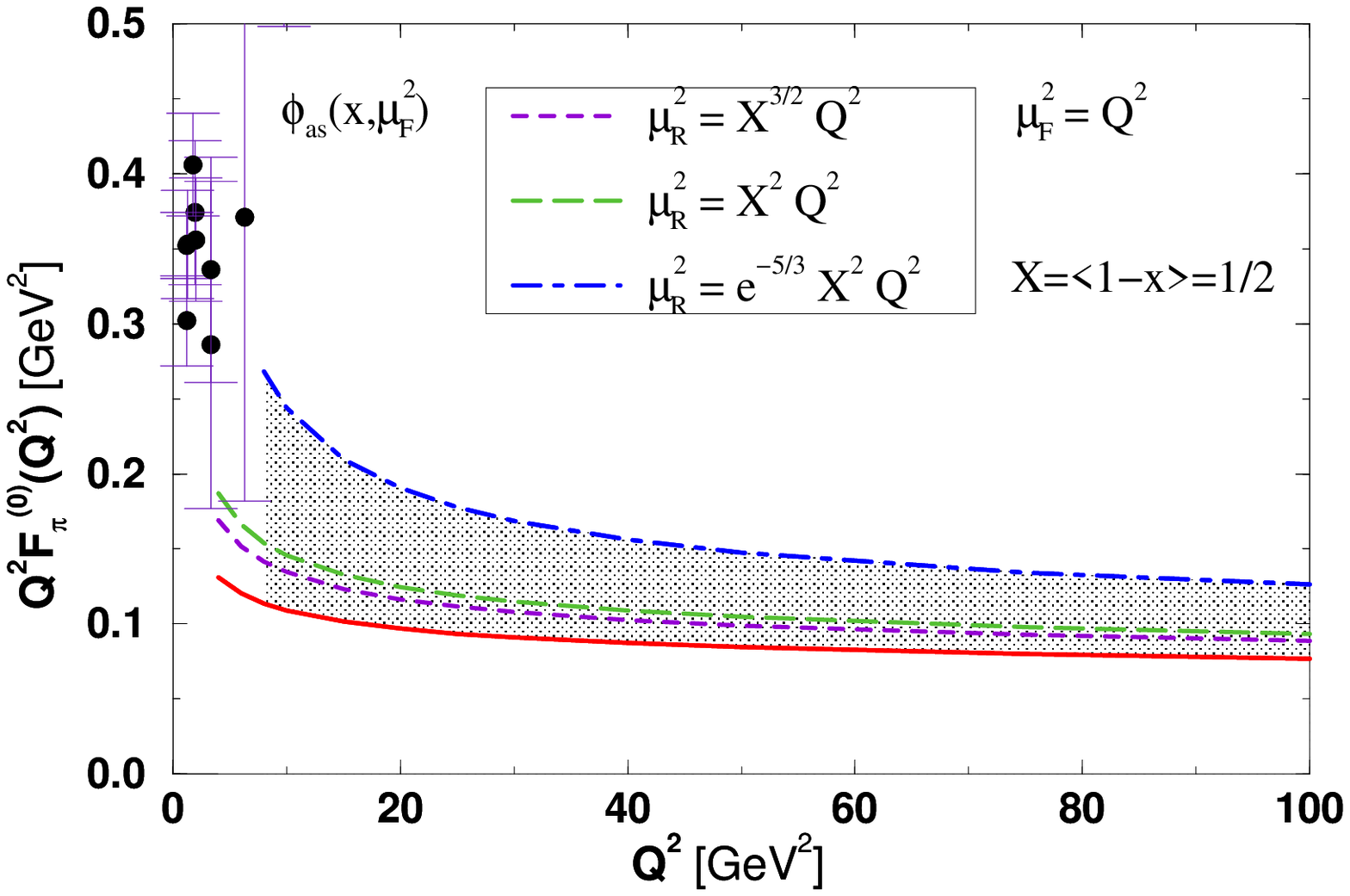,height=8cm,width=11cm,silent=}}
\caption{The numerical results for $Q^2 F_{\pi}^{(0)}(Q^2)$ obtained with
  the $\phi_{as}(x,\nIR)$ distribution amplitude
  and the choices of $\nUV$ given by
  Eqs. \protect\req{eq:nR},
  with $\nIR=Q^2$ and $\left< \x \right>_{as}=1/2$. 
  The solid curve (included for comparison) 
  corresponds to the case $\nUV=\nIR=Q^2$
  considered in the preceding subsection.
  The shaded area denotes the range of the LO prediction
  $Q^2 F_{\pi}^{(0)}(Q^2)$ for 
  $\nUV/Q^2 \in \left[e^{-5/3} \left< \x \right>^2, 1 \right]$.
  }
\label{f:asAb}
\end{figure}
\begin{figure}
\centerline{\epsfig{file=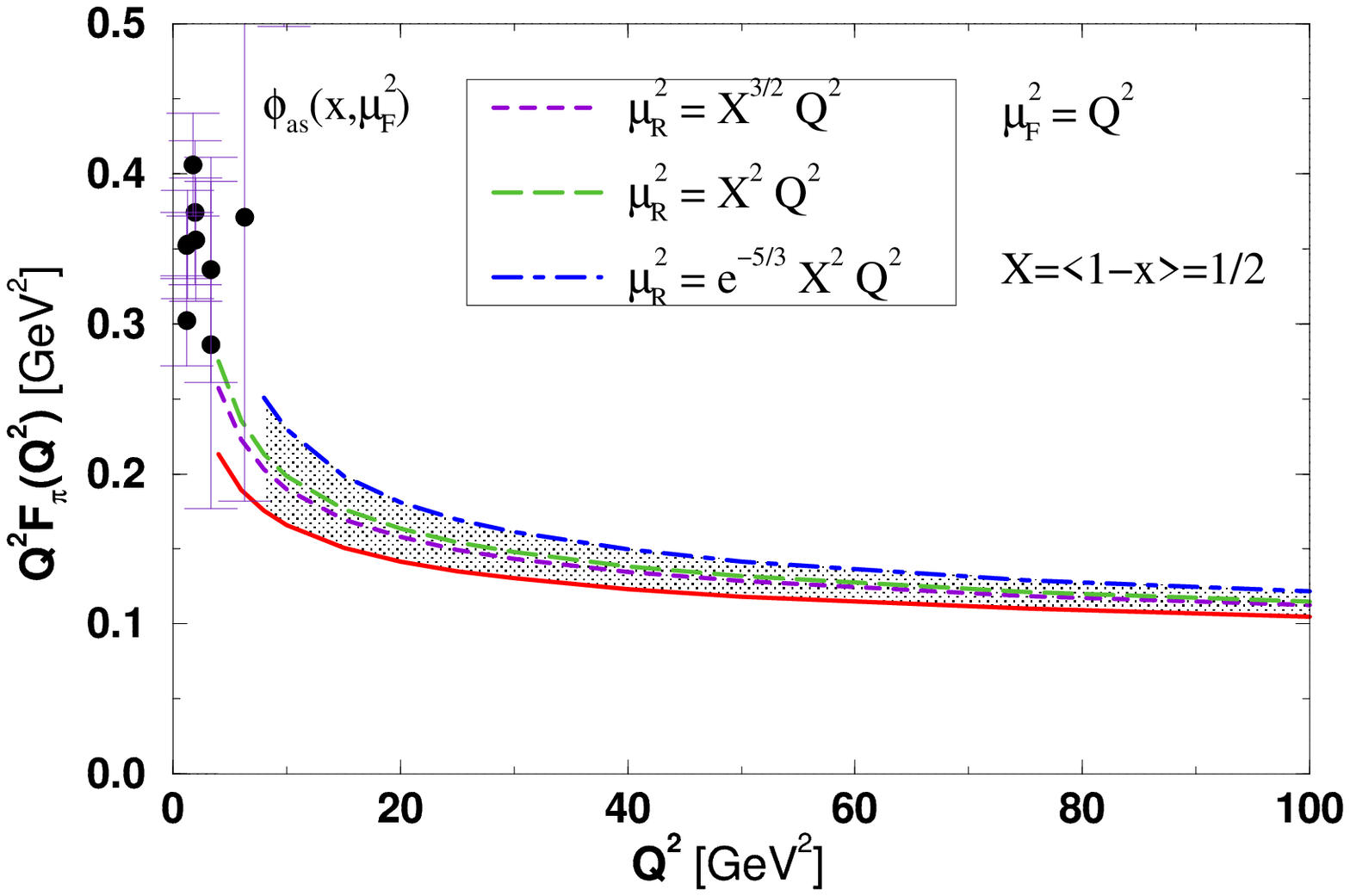,height=8cm,width=11cm,silent=}}
\caption{Leading-twist NLO QCD results for $Q^2 F_{\pi}(Q^2)$ obtained with
  the $\phi_{as}(x,\nIR)$ distribution amplitude
  and the choices of $\nUV$ given by
  Eqs. \protect\req{eq:nR},
  with $\nIR=Q^2$ and $\left< \x \right>_{as}=1/2$. 
  The solid curve (included for comparison) 
  corresponds to the case $\nUV=\nIR=Q^2$
  considered in the preceding subsection.
  The shaded area denotes the range of the 
  total NLO prediction $Q^2 F_{\pi}(Q^2)$ 
  where the upper limit corresponds to $\nUV=\mu_{PMS}^2$
  obtained from \protect\req{eq:PMS}.
  }
\label{f:ASbA}
\centerline{\epsfig{file=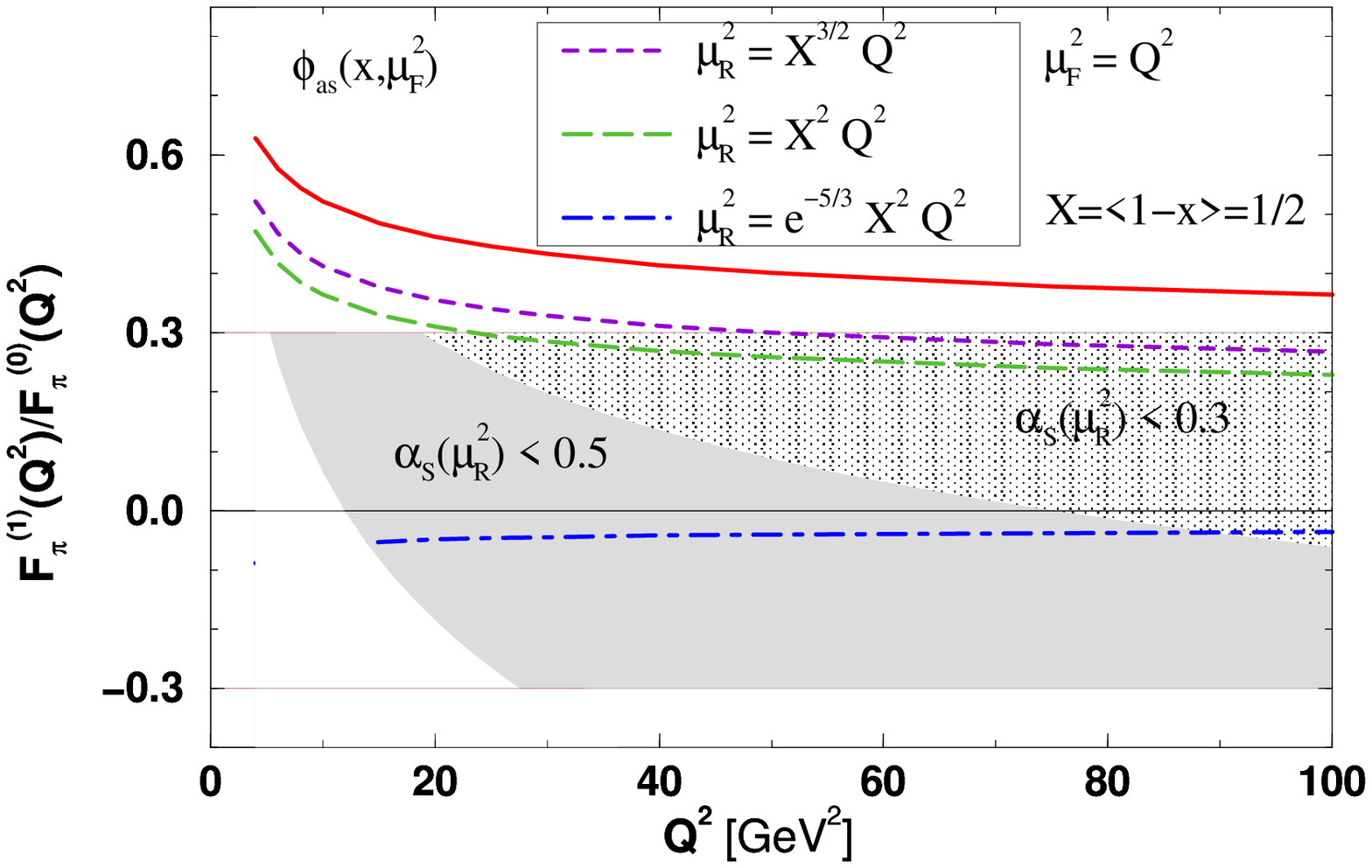,height=8cm,width=11cm,silent=}}
\caption{The ratio $F_{\pi}^{(1)}(Q^2)/F_{\pi}^{(0)}(Q^2)$ obtained with
  the $\phi_{as}(x,\nIR)$ distribution amplitude 
  and the choices of $\nUV$ given by
  Eqs. \protect\req{eq:nR},
  with $\nIR=Q^2$ and $\left< \x \right>_{as}=1/2$. 
  The solid curve (included for comparison) 
  is for the result corresponding  to $\nUV=\nIR=Q^2$
  obtained in the preceding subsection, and
the shaded area denotes the region of the predictions
which correspond to 
$\alpha_S(\nUV)\!<\!0.5$ ($\alpha_S(\nUV)\!<\!0.3$),
 while $|F_{\pi}^{(1)}(Q^2)/F_{\pi}^{(0)}(Q^2)| < 0.3$. 
  }
\label{f:ratioASbA}
\end{figure}
\begin{figure}
\centerline{\epsfig{file=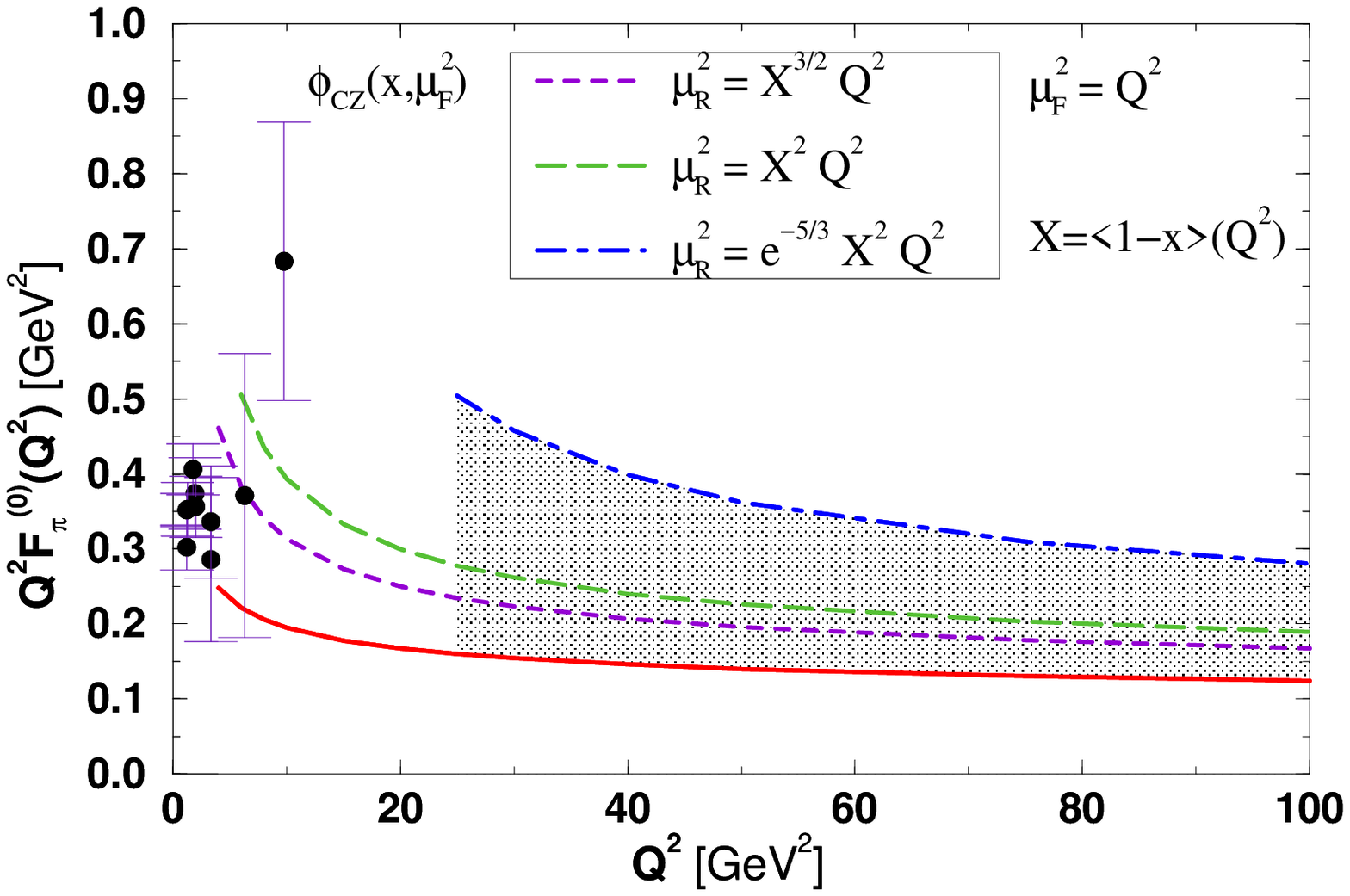,height=8cm,width=11cm,silent=}}
\caption{The numerical results for $Q^2 F_{\pi}^{(0)}(Q^2)$ obtained with
  the $\phi_{CZ}(x,\nIR)$ distribution amplitude
  and the choices of $\nUV$ given by
  Eqs. \protect\req{eq:nR},
  with $\nIR=Q^2$ and $\left< \x \right>_{CZ}(Q^2)$ 
         calculated according to \protect\req{eq:sx3}.
  The solid curve (included for comparison) 
  corresponds to the case $\nUV=\nIR=Q^2$
  considered in the preceding subsection.
  The shaded area denotes the range of the LO prediction
  $Q^2 F_{\pi}^{(0)}(Q^2)$ for 
  $\nUV/Q^2 \in \left[e^{-5/3} \left< \x \right>^2, 1 \right]$.
  }
\label{f:czAcQ}
\end{figure}
\begin{figure}
\centerline{\epsfig{file=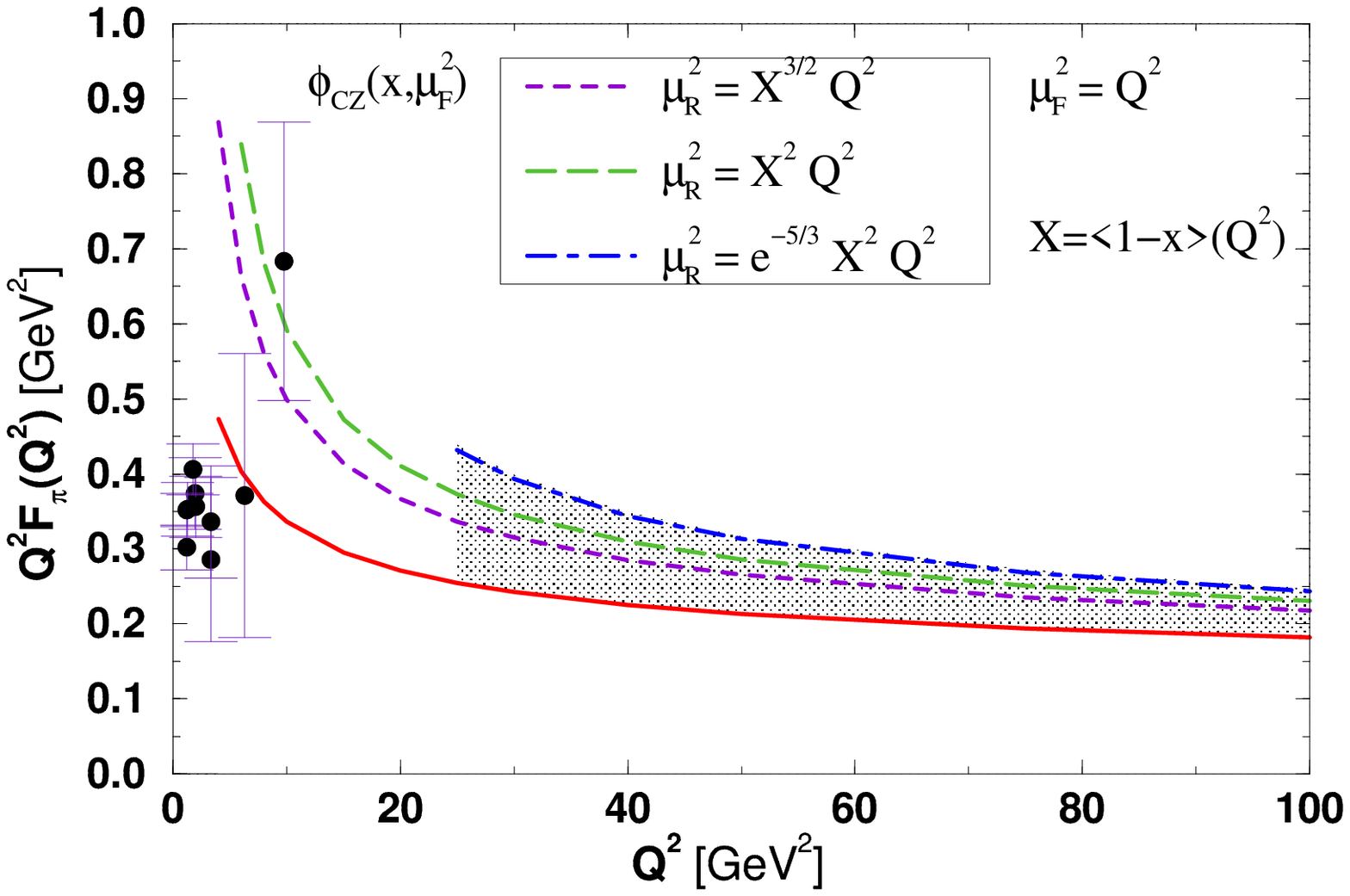,height=8cm,width=11cm,silent=}}
\caption{Leading-twist NLO QCD results for $Q^2 F_{\pi}(Q^2)$ obtained with
  the $\phi_{CZ}(x,\nIR)$ distribution amplitude
  and the choices of $\nUV$ given by
  Eqs. \protect\req{eq:nR},
  with $\nIR=Q^2$ and $\left< \x \right>_{CZ}(Q^2)$ 
         calculated according to \protect\req{eq:sx3}.
  The solid curve (included for comparison)
  corresponds to the case $\nUV=\nIR=Q^2$
  considered in the preceding subsection.
  The shaded area denotes the range of the 
  total NLO prediction $Q^2 F_{\pi}(Q^2)$ 
  where the upper limit corresponds to $\nUV=\mu_{PMS}^2$
  obtained from \protect\req{eq:PMS}.
  }
\label{f:CZcQA}
\centerline{\epsfig{file=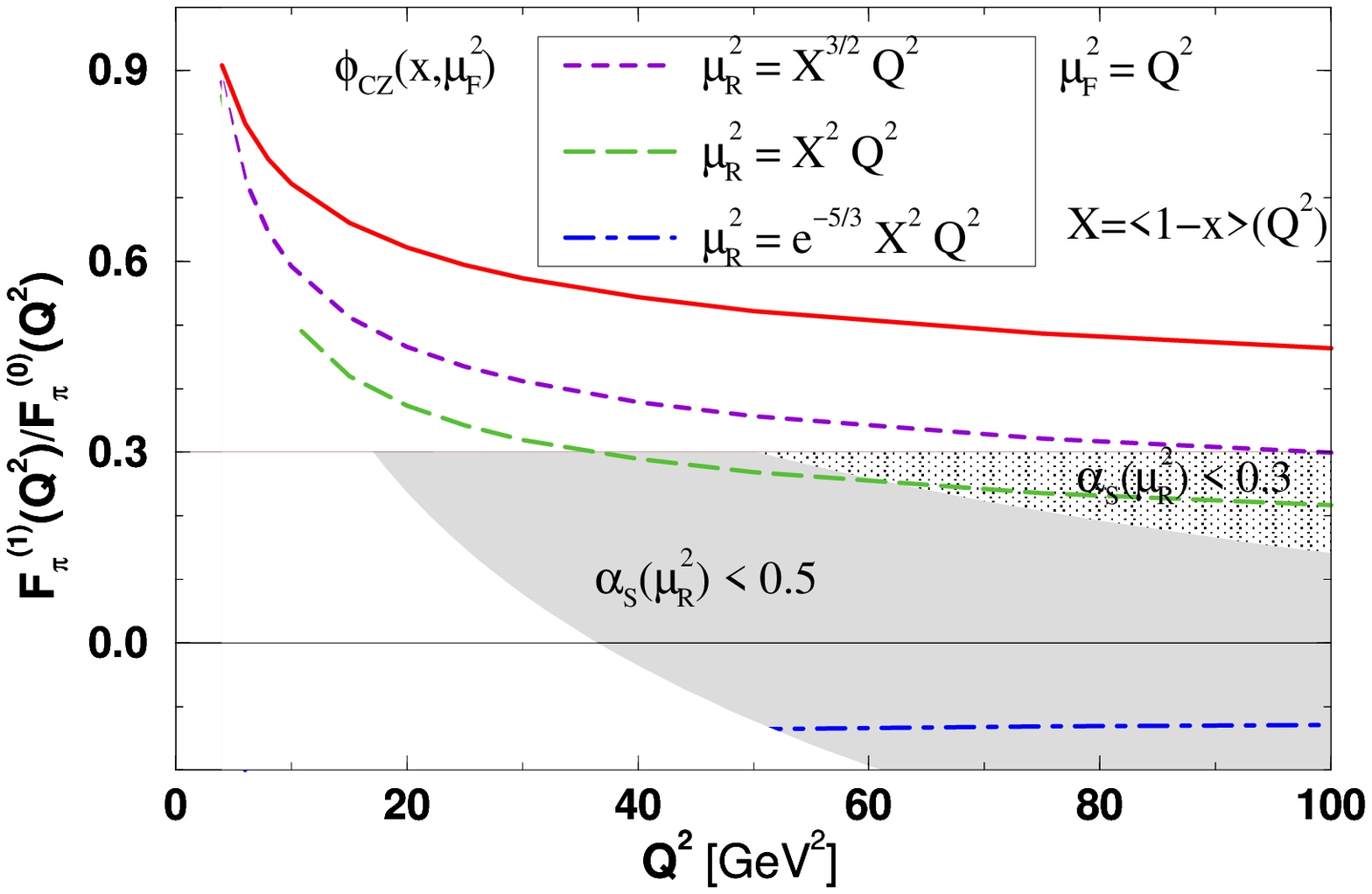,height=8cm,width=11cm,silent=}}
\caption{The ratio $F_{\pi}^{(1)}(Q^2)/F_{\pi}^{(0)}(Q^2)$ obtained with
  the $\phi_{CZ}(x,\nIR)$ distribution amplitude 
  and the choices of $\nUV$ given by
  Eqs. \protect\req{eq:nR},
  with $\nIR=Q^2$ and $\left< \x \right>_{CZ}(Q^2)$ 
         calculated according to \protect\req{eq:sx3}.
  The solid curve (included for comparison) 
  is for the result corresponding  to $\nUV=\nIR=Q^2$
  obtained in the preceding subsection, and
the shaded area denotes the region of the predictions
which correspond to 
$\alpha_S(\nUV)\!<\!0.5$ ($\alpha_S(\nUV)\!<\!0.3$),
 while $|F_{\pi}^{(1)}(Q^2)/F_{\pi}^{(0)}(Q^2)| < 0.3$. 
  }
\label{f:ratioCZcQA}
\end{figure}

The three specific physically motivated choices of ${\mu}_R$, 
given by \req{eq:nR},
can be conveniently written as
\Bml
\label{eq:aQ2}
\Beq
  \nUV=a \, Q^2,
\Eeq
where
\Beq
    a \in
         \{
{\left < \x \right >}^{3/2},
{\left < \x \right >}^{2},
e^{-5/3} {\left < \x \right >}^{2}
         \} \, .
\Eeq
\Eml

Our complete leading-twist NLO predictions for the pion form factor
calculated with the
${\phi}_{as}(x,\nIR)$
distribution and with the three specific values of the renormalization
scale ${\mu}_R$ are summarized in 
Figs. \ref{f:ASbA} and \ref{f:ratioASbA}, showing the
results for $Q^2F_{\pi}(Q^2)$
and the ratio $F_{\pi}^{(1)}(Q^2)/ F_{\pi}^{(0)}(Q^2)$,
respectively.
For the purpose of the discussion of the effects of the inclusion of the NLO
corrections on the LO predictions,
we have also included 
Fig. \ref{f:asAb} 
showing the LO predictions for the three different
values of ${\mu}_R$
given in Eq. \req{eq:aQ2}.
The corresponding results but obtained assuming the
${\phi}_{CZ}(x,\nIR)$
distribution are displayed in 
Figs. \ref{f:czAcQ}, \ref{f:CZcQA}, and \ref{f:ratioCZcQA}.

The solid curves in Figs. \ref{f:asAb} - \ref{f:ratioCZcQA}
correspond to the results with $\nUV=\nIR=Q^2$
obtained in the preceding subsection,
are included for comparison.
The other curves refer to the choices of $\nUV$ given by
\req{eq:aQ2}.

Some general comments concerning the results presented above are in order.
First of all, it is interesting to note that the predictions based on the
${\phi}_{as}(x,\nIR)$ and ${\phi}_{CZ}(x,\nIR)$
distributions amplitudes almost share the qualitative features, 
although these two distributions are quite different in shape.
Next, by looking at 
Figs. \ref{f:ASbA} and \ref{f:CZcQA}, one notices that
the full NLO result for $Q^2F_{\pi}(Q^2)$,
shows a very weak dependence on the value of $\nUV$,
and that it increases with decreasing $\nUV$.
As far as the ratio $F_{\pi}^{(1)}(Q^2)/ F_{\pi}^{(0)}(Q^2)$
is concerned, as evident from 
Figs. \ref{f:ratioASbA} and \ref{f:ratioCZcQA},
the situation is quite different.
It is rather sensitive to the variation of
$\nUV$, and decreases as $\nUV$ decreases.
The choice $\nUV=Q^2$,
represented by solid curves,
when compared with the other possibilities considered, leads
to the lowest value for $Q^2F_{\pi}(Q^2)$,
and to the highest value for the ratio
$F_{\pi}^{(1)}(Q^2)/ F_{\pi}^{(0)}(Q^2)$.
In contrast to that,
the choice of the BLM scale,
represented by the dashed-dotted curve,
leads to somewhat higher values of
$Q^2F_{\pi}(Q^2)$,
but to considerably lower values of the ratio
$F_{\pi}^{(1)}(Q^2)/ F_{\pi}^{(0)}(Q^2)$.

Adopting the previously stated criteria,
we now comment on the reliability of the NLO predictions for
$Q^2F_{\pi}(Q^2)$
displayed in Figs. \ref{f:ASbA} and \ref{f:CZcQA}.
Imposing the requirements
$\left| F_{\pi}^{(1)}(Q^2)/ F_{\pi}^{(0)}(Q^2) \right|<0.3$,
and ${\alpha}_S (\nUV)<0.3$,
we find from Fig. \ref{f:ratioASbA} that for the
${\phi}_{as}(x,{\mu}_F^2)$ distribution
the results corresponding to
$a={\left < \x \right >}^{3/2}$,
${\left < \x \right >}^{2}$,
and
$e^{-5/3} {\left < \x \right >}^{2}$,
become reliable for the momentum transfer
$Q^2>50$, $25$, and $90$ GeV$^2$, respectively.
It might be that the requirement
${\alpha}_S (\nUV)<0.3$
is too stringent.
Thus, by relaxing it and taking
${\alpha}_S (\nUV)<0.5$
instead,
we find that the results corresponding to
$a={\left < \x \right >}^{3/2}$,
${\left < \x \right >}^{2}$,
and
$e^{-5/3} {\left < \x \right >}^{2}$,
become reliable for the momentum transfer
$Q^2>50$, $25$, and $15$ GeV$^2$, respectively.

Applying the same criteria to the case of the
${\phi}_{CZ}(x,\nUV)$
distribution, we find from Fig. \ref{f:ratioCZcQA}
that the results for $Q^2F_{\pi}(Q^2)$, 
shown in Fig. \ref{f:CZcQA}, obtained with
$a={\left < \x \right >}^{3/2}$,
${\left < \x \right >}^{2}$,
and
$e^{-5/3} {\left < \x \right >}^{2}$,
become reliable for the momentum transfer
$Q^2>95$, $60$, and $300$ GeV$^2$, respectively,
if we demand that
${\alpha}_S (\nUV)<0.3$,
and for 
$Q^2>95$, $35$, and $50$ GeV$^2$,
respectively, if
${\alpha}_S (\nUV)<0.5$
is regarded as a stringent enough requirement.

Summarizing the above, one can say that contrary to the rather high value
of $Q^2 \approx 500$ GeV$^2$
required to obtain a reliable prediction assuming the
${\phi}_{as}(x, \nUV)$
distribution with
$\nUV=Q^2$, one finds that by choosing the renormalization scale
determined by the dynamics of the pion rescattering process, the size of
the NLO corrections is significantly reduced and reliable predictions are
obtained  at considerably lower values of $Q^2$, namely, for $Q^2<100$ GeV$^2$.
The same conclusion also applies to the results obtained with the
${\phi}_{CZ}(x,\nUV)$
distribution, for which choosing the renormalization scale related to the
virtuality of the particles in the parton subprocess lowers the bound
of the reliability of the results from
$Q^2 \approx 2400$ GeV$^2$ 
to
$Q^2 < 100$ GeV$^2$. 

\subsubsection{Theoretical uncertainty of the NLO predictions
              related to the renormalization scale ambiguity}

Unfortunately, at present we do not have at our disposal any
absolutely reliable method of determining the ``optimal'' (or ``correct'')
value of the renormalization scale
for any particular order of PQCD.
Our ignorance concerning the ``optimal'' value for
${\mu}_R$ implies
that any particular choice of this scale leads to an intrinsic 
theoretical uncertainty  (error) of the perturbative results.
Therefore, the NLO results displayed  in 
Figs. \ref{f:ASbA} and \ref{f:CZcQA},
being obtained with the four 
singled-out values of the renormalization scale ${\mu}_R$, contain
theoretical uncertainty. 
In what follows we try to estimate this uncertainty.
In order to do that, we have to make some assumptions, regarding
the range of the renormalization scale ambiguity.

As already mentioned in addition to the BLM method, two more
renormalization scale-settings have been proposed:
the FAC and PMS methods.
All three of them
are somewhat ad hoc and have no strong justification.
Nevertheless, the principles underlying these methods are plausible, 
so that they give
us at least a range of scale ${\mu}_R$ which should be considered.

Let $\mu_{FAC}$, $\mu_{PMS}$, and $\mu_{BLM}$ designate
the scales determined by the BLM, PMS, and FAC scale setting methods,
respectively.

According to 
the FAC procedure,
the scale ${\mu}_R$ is determined by the requirement that
the NLO coefficient in the perturbative expansion of $F_{\pi}(Q^2)$ vanishes,
which, in our case, effectively reduces to solving the equation
\Beq
F_{\pi}^{(1)}(Q^2,\nUV=\mu_{FAC}^2)=0.
\label{eq:FAC}
\Eeq

On the other hand,
in the PMS procedure, 
one chooses the renormalization scale
${\mu}_R$ at the stationary point of the truncated perturbative series for
$F_{\pi}(Q^2)$. Operationally, this amounts to
\Beq
\frac {dF_{\pi}(Q^2,\nUV)} {d\nUV} 
         \left|_{\displaystyle\nUV=\displaystyle\mu_{PMS}^2}
	 \right.= 0.
\label{eq:PMS}
\Eeq

The BLM-determined scale is given by
\Beq
      \mu_{BLM}^2=e^{-5/3} \left< \x \right>^2 Q^2 \, .
\label{eq:BLM}
\Eeq

The explicit expressions for $F_{\pi}(Q^2)$ and $F_{\pi}^{(1)}(Q^2)$
are given by (\ref{eq:Q2Fpi0}-\ref{eq:Q2Fpi})
with $\nIR=Q^2$.
By solving Eqs. \req{eq:FAC} and \req{eq:PMS},
and taking \req{eq:BLM} into account 
we find that 
for the ${\phi}_{as}(x,{\mu}_F^2)$ distribution:
\Bml
\label{eq:ASscales}
\Beqa
& & \mu_{BLM}^2 = Q^2/21 \, ,
   \\
& & \mu_{PMS}^2 \approx Q^2/18 \, ,
  \label{eq:ASpms} \\
& & \mu_{FAC}^2 \approx Q^2/18 
 \, ,
\Eeqa
\Eml
while, for the
${\phi}_{CZ}(x,{\mu}_F^2)$ distribution:
\Bml
\label{eq:CZscales}
\Beqa
Q^2/79 < \mu_{BLM}^2 < Q^2/77 
   \, ,  \\
Q^2/59 < \mu_{PMS}^2 < Q^2/54 
   \, ,  \label{eq:CZpms} \\
Q^2/53 < \mu_{FAC}^2 < Q^2/48 
\, , 
\Eeqa
\Eml
for $50$ GeV$^2 < Q^2 < 100$ GeV$^2$.
Therefore, we find that for both distributions
\Beq
   \mu_{BLM}^2 < \mu_{PMS}^2 < \mu_{FAC}^2
     \, .
\Eeq

The FAC, PMS, and BLM scales,
as evident from \req{eq:ASscales} and \req{eq:CZscales},
are very close to each other,
and the curves corresponding to the NLO prediction for 
$Q^2 F_{\pi}(Q^2)$ obtained with the FAC and PMS scales practically
coincide 
with the dashed-dotted curves in Figs. 
\ref{f:ASbA} and \ref{f:CZcQA} corresponding to the BLM scale.

If the renormalization scale is interpreted as a ``typical'' scale
of virtual momenta in the corresponding Feynman diagrams, then, 
despite the fact that we do not know what the ``best'' value
of this scale is, it is 
(based on physical grounds and on the above considerations) 
reasonable 
to assume that it belongs to the interval
ranging from ${\mu}_{BLM}^2$ to $Q^2$. 
Namely, 
${\mu}_{BLM}^2$, being the lowest of the above considered scales
and of the order of
$Q^2/21$ and $Q^2/80$  
for the
${\phi}_{as}(x,{\mu}_F^2)$
and
${\phi}_{CZ}(x,{\mu}_F^2)$
distributions, respectively,
is a low enough to serve as the 
lower limit of the renormalization scale ambiguity interval. 
On the other hand,
$Q^2$ is (too) high a scale and 
as such can safely be used for 
the upper limit of the same interval.

In the following, then, instead of using any singled-out value,
we vary the renormalization scale in the form
\Bml
\label{eq:interval}
\Beq
   \nUV= a \, Q^2
   \, ,
\Eeq
where $a$ is a continuous parameter
\Beq
    a \in [ e^{-5/3} \left< \x \right>^2, 1 ]
    \, .
\Eeq
\Eml
Doing this will enable us to draw some qualitative conclusions concerning,
first, the theoretical uncertainty related to the renormalization
scale ambiguity, and, second, the effects that the inclusion of the NLO
corrections has on the LO predictions.

The LO result for the pion form factor
is a monotonous function of the renormalization scale $\mu_R$.
Namely, all of the $\mu_R$ dependence of the LO prediction
$Q^2 F_{\pi}^{(0)}(Q^2)$, 
as it is seen from \req{eq:Q2Fpi0},
is contained in the strong coupling constant
$\alpha_S(\nUV)$.
Thus, in accordance with \req{eq:alphas},
as $\mu_R$ decreases the LO result increases, and
it increases without bound.
In contrast to the LO,
the NLO contribution
$Q^2 F_{\pi}^{(1)}(Q^2)$, as evident from the explicit expression
given in \req{eq:Q2Fpi1a}, decreases (becomes more negative)
with decreasing $\mu_R$.
Upon adding up the LO and NLO contributions, 
we find that the full NLO result,
as a function of $\mu_R$ stabilizes and reaches
a maximum value for $\nUV=\mu_{PMS}^2$. 
The values of the scale $\mu_{PMS}$ are, for the
$\phi_{as}(x,Q^2)$ and $\phi_{CZ}(x,Q^2)$ distributions given by
\req{eq:ASpms} and \req{eq:CZpms}, respectively.

If the renormalization scale continuously changes in the interval
defined by \req{eq:interval}, 
we find that the curves representing the LO and NLO results for
$Q^2 F_{\pi}(Q^2)$ 
fill out the shaded regions in 
Figs. \ref{f:asAb} and \ref{f:ASbA},
for the $\phi_{as}(x,Q^2)$, and  in
Figs. \ref{f:czAcQ} and \ref{f:CZcQA}
for the $\phi_{CZ}(x,Q^2)$ distribution.

Next we turn to discuss the intrinsic theoretical uncertainty
of the NLO prediction 
related to the renormalization scale ambiguity.
Regarding this uncertainty, there is in fact no consensus on how to estimate
it, or how to identify what the central value of $Q^2 F_{\pi}(Q^2)$
should be.

Nevertheless, the simplest but still a good measure of this
uncertainty is the quantity
\Beq
{\Delta} F_{\pi}(Q^2, {\mu}_{min}^2,{\mu}_{max}^2)=
F_{\pi}(Q^2,{\mu}_{min}^2)-F_{\pi}(Q^2,{\mu}_{max}^2)
 \, ,
\label{eq:uncertainty}
\Eeq
i.e., the difference of the results for $F_{\pi}(Q^2)$ corresponding to the
lower and the upper limit of the renormalization scale ambiguity interval
\req{eq:interval}. 
This quantity, therefore, for a given value of $Q^2$, represents the 
``width'' of the shaded regions in 
Figs. \ref{f:asAb}, \ref{f:ASbA}, \ref{f:czAcQ}, and \ref{f:CZcQA}.
In this sense, the shaded regions in these figures
limited by the curves corresponding to $\nUV=Q^2$
and $\nUV=\mu_{BLM}^2$ essentially determine the theoretical 
accuracy allowed by the LO and NLO calculation.

A glance at 
Figs. \ref{f:asAb} and \ref{f:ASbA},
displaying the LO and NLO predictions for
$Q^2 F_{\pi}(Q^2)$ calculated with the $\phi_{as}(x,Q^2)$
distribution, reveals that, compared to the LO, the NLO results exibit a
much smaller renormalization
scale dependence.
The same holds true for the predictions 
depicted in
Figs. \ref{f:czAcQ} and \ref{f:CZcQA},
based on the $\phi_{CZ}(x,Q^2)$ distribution.

To be more quantitative, we thus find that,
if, for the $\phi_{as}(x,Q^2)$ distribution,
at $Q^2=50$ GeV$^2$ ($100$ GeV$^2$), 
instead of
${\mu}_R^2=Q^2$
one takes
${\mu}_R^2 = \mu_{BLM}^2 = Q^2/21$, 
the LO result (Fig. \ref{f:asAb}) increases by
$75\%$ ($64\%$),
whereas the NLO result (Fig. \ref{f:ASbA})
increases by $20\%$ ($16\%$).
Analogously, for the same values of $Q^2$,
but for the $\phi_{CZ}(x,Q^2)$ distribution we
find that taking 
${\mu}_R^2 =\mu_{BLM}^2 \approx Q^2/79$ 
instead $\nUV=Q^2$ the LO result increases by
$158\%$ ($125\%$), 
while the  NLO result increases by $47\%$ ($34\%$). 

Therefore, the NLO corrections improve the situation 
because the terms in the NLO hard-scattering amplitude arise that
cancel part of the scale dependence of the LO result.

It should be pointed out that our estimate of the
renormalization scale ambiguity interval given in
\req{eq:interval} is very conservative, 
overestimating the theoretical uncertainty of the calculated NLO
predictions.
Namely, one could, almost at no risk replace $Q^2$ by
$\left< \x \right>^{3/2} Q^2$ as the upper limit of the 
interval. 
If this is done, the dotted rather than solid curves
would then provide the lower bound of the shaded regions in
Figs. \ref{f:ASbA} and \ref{f:CZcQA}.
Then, for $Q^2 \geq 50$ GeV$^2$ 
theoretical uncertainty of the NLO result for 
$Q^2 F_{\pi}(Q^2)$ turns out to be less than
$5\%$ for the $\phi_{as}(x,Q^2)$ 
and $8\%$ for the $\phi_{CZ}(x,Q^2)$ distribution.

Before closing this subsection, a remark is appropriate.
If the shaded areas in Figs. 
\ref{f:ASbA} and \ref{f:CZcQA}
are displayed in the same figure they 
would not overlap.
This implies that an unambiguous discrimination between
the $\phi_{as}(x,Q^2)$ and $\phi_{CZ}(x,Q^2)$
distributions is possible, as soon as the data
extending to higher values of $Q^2$ are obtained.

Based on the above considerations, 
we may conclude that the inclusion of the NLO corrections 
stabilizes the LO prediction for the pion form factor
by considerably reducing the intrinsic theoretical uncertainty related 
to the renormalization scale ambiguity.
This uncertainty for both distributions turns out to be
of the order of a few percent. 


\section{Summary and conclusions}

In this paper we have presented the results of a complete leading$-$twist
NLO QCD analysis of the spacelike electromagnetic form factor of the pion
at large momentum transfer.

To clarify the discrepancies in the analytical expression for the
hard$-$scattering amplitude present in previous calculations,
we have carefully recalculated  the one$-$loop Feynman diagrams shown in
Fig. \ref{f:oldi}. 
Working in the $\overline {\rm MS}$ renormalization scheme and
employing the dimensional regularization method to treat all divergences
(UV, IR, and collinear), we have obtained results which are in 
agreement with those of Refs. \cite{FiG81} 
(up to the typographical errors listed in \cite{BraT87})
and \cite{Sa82}.

As nonperturbative input at the reference momentum scale of 0.5 GeV, we
have used the four available pion distribution amplitudes
defined by Eq. \req{eq:phix} and plotted in Fig. \ref{f:xDA}:
the asymptotic distribution ${\phi}_{as}(x,{{\mu}_F}^2)$ and the three QCD
sum$-$rule inspired distributions
${\phi}_{CZ}(x,{{\mu}_F}^2)$,
${\phi}_{P2}(x,{{\mu}_F}^2)$, and
${\phi}_{P3}(x,{{\mu}_F}^2)$.
The NLO evolution of these distributions has been determined using
the formalism developed in Ref. \cite{Mu94etc}.

By convoluting, according to Eq. \req{eq:piffcf}, 
the hard-scattering amplitude with
the pion distribution amplitude, both calculated in the NLO approximation,
we have obtained the NLO numerical predictions for the pion form factor,
for the four candidate distributions, and for several different choices of the
renormalization and factorization scales, ${\mu}_R$ and ${\mu}_F$.
All the predictions have been obtained assuming $n_f=3$ and
${\Lambda}_{\overline {MS}}$=0.2 GeV.

We have first used the most simple choice of the scales
where ${{\mu}_R}^2={{\mu}_F}^2=Q^2$. The results are
summarized in Figs. \ref{f:Aa} and \ref{f:ratioAa} 
and Tables \ref{t:rASaA}--\ref{t:rP3aA}.
Although the predictions for the pion form factor $Q^2 F_{\pi}(Q^2)$
obtained with the four candidate distributions differ considerably,
the lack of reliable experimental data, especially in the higher
$Q^2$ region, does not allow us to confirm or discard any of them.
The size of the NLO corrections and the size of the running coupling constant
at given $Q^2$ can be used as indicators of the reliability of
the perturbative treatment. 
Our numerical results based on the asymptotic distribution amplitude
${\phi}_{as}(x,{{\mu}_F}^2)$
differ from those of Ref. \cite{FiG81} (the difference is due to
the different value of $\Lambda_{\overline{MS}}$). 
Thus, in contrast to Ref. \cite{FiG81}, where it
was concluded that ``reliable perturbative predictions can not be made
until momentum transfers Q of about 100 GeV are reached'',
we have found that reliable predictions can already be made at 
momentum transfers of the order of 25 GeV.
It has been shown that the inclusion of the (NLO) evolutional
corrections only slightly influences  the NLO prediction
obtained assuming the $\phi_{as}(x,\nIR)$ distribution.
On the other hand, for the case of the end-point concentrated
distributions, the evolutional corrections, both the LO and the NLO, 
are important. 
For the $\nUV=\nIR=Q^2$ choice of scales, the NLO corrections based on the
$\phi_{as}(x,\nIR)$, $\phi_{CZ}(x,\nIR)$ 
$\phi_{P2}(x,\nIR)$, and $\phi_{P3}(x,\nIR)$ distributions,
are large,
which implies that one must demand
that the momentum transfer $Q$ be considerably larger than $10$ GeV
before the corresponding results become reliable.

In order to reduce the size of the NLO corrections
and to examine the extent to which the NLO predictions for the pion
form factor depend on the scales
${\mu}_R$ and ${\mu}_F$, in addition to the simplest choice
${{\mu}_R}^2={{\mu}_F}^2=Q^2$
(which certainly is not best suited for the process of interest),
we have also considered
the choices of ${\mu}_R$ and ${\mu}_F$ given by 
Eqs. \req{eq:nR} and \req{eq:snIR}, respectively.
The results appear to be insensitive to the choice of
$\mu_F$ considered. 
Using alternative choices for $\mu_R$, and modeling the
pion with the
${\phi}_{as}(x,Q^2)$ and
${\phi}_{CZ}(x,Q^2)$ distribution amplitudes,  
leads to the predictions shown in Figs. \ref{f:ASbA},
\ref{f:ratioASbA}, \ref{f:CZcQA}, and \ref{f:ratioCZcQA}. 
The ${\phi}_{P2}(x,{{\mu}_F}^2)$ and 
${\phi}_{P3}(x,{{\mu}_F}^2)$ distributions are not separately considered,
since the corresponding results are very similar to those obtained with 
${\phi}_{CZ}(x,{{\mu}_F}^2)$.

For a given distribution amplitude, the values of the pion form factor
$Q^2F_{\pi}(Q^2)$
are rather insensitive to various choices of the scales
${\mu}_R$. 
This is evident from Figs. \ref{f:ASbA} and
\ref{f:CZcQA}, 
and is a reflection of the stabilizing effect that the inclusion of the
NLO corrections has on the LO predictions.
On the other hand, the
ratio of the NLO corrections to the LO prediction,
$F_{\pi}^{(1)}(Q^2)/F_{\pi}^{(0)}(Q^2)$ 
is very sensitive to the values of ${\mu}_R$,
as can be seen from Figs. \ref{f:ratioASbA} and \ref{f:ratioCZcQA}.
Requiring this ratio to be less than $0.3$ and taking the less stringent 
condition on the value of the strong coupling $\alpha_S(\nUV) < 0.5$,  
we find 
that the predictions 
can be considered reliable for the momentum transfer  
$Q < 10$ GeV, 
provided the renormalization scale related to the average virtuality 
of the particles in the parton subprocess or given by the BLM scale, 
is used.

Given the fact that we do not know what the ``optimal'' value
of the renormalization scale is, choosing any particular value
for this scale introduces a theoretical uncertainty in the NLO
predictions.
Based on a reasonable guess of the renormalization
scale ambiguity interval, we have estimated this uncertainty
to be  less than $10\%$.

The difference between the absolute predictions based on the 
$\phi_{as}(x,\nIR)$, 
$\phi_{CZ}(x,\nIR)$ ($\phi_{P2}(x,\nIR)$), and $\phi_{P3}(x,\nIR)$ 
distributions is large enough to allow 
an unambiguous experimental discrimination between them, as soon as 
the data extending to higher values of $Q$ become available. 

In conclusion, the results of the complete leading-twist NLO QCD analysis,
which has been carried out in this paper, show that reliable 
perturbative predictions for the pion electromagnetic form factor with 
all the four distribution amplitudes considered can already 
be made at a momentum
transfer $Q<10$ GeV. 
The theoretical uncertainty related to the renormalization scale
ambiguity, which constitutes a reasonable range of physical
values, has been shown to be less than $10\%$.
To check our predictions and to choose 
between the distribution amplitudes, it is necessary that 
experimental data at higher values of $Q^2$ are obtained.

\acknowledgments

  The authors would like to thank A.~V. Radyushkin
  for pointing out an error present in the original version of the manuscript,
  and P. Kroll for useful suggestions.
  This work was supported by the Ministry of Science and Technology
  of the Republic of Croatia under Contract No. 00980102.

\end{document}